\documentstyle[psfig]{mn}

\def\eta{et al.}
\def\ma{~\raise.5ex\hbox{$>$}\kern-.8em\lower 1mm\hbox{$\sim$}~}
\def\ergs{erg cm$^{\rm -2}$ s$^{\rm -1}$}

\title{X-ray properties of the Parkes sample of flat-spectrum radio sources:
dust in radio-loud quasars?}

\author[J.~Siebert et al.]{J.~Siebert,$^1$ W.~Brinkmann,$^1$ 
M.J.~Drinkwater,$^2$ W.~Yuan,$^1$ P.J.Francis,$^3$ 
\and B.A.~Peterson$^3$ and R.L.~Webster$^4$\\ 
$^1$ Max-Planck-Institut f\"ur extraterrestrische Physik,
 Giessenbachstrasse, D-85740 Garching, Germany\\ 
$^2$ School of Physics, University of New South Wales, Sydney 2052, Australia\\
$^3$ Mt. Stromlo \& Siding Spring Observatories, Australian National 
 University, Weston Creek, A.C.T. 2611, Australia\\
$^4$ School of Physics, University of Melbourne, Parkville, Victoria 3052,
Australia}

\date{Received; accepted}
\pubyear{1998}

\begin{document}

\maketitle

\begin{abstract}

We investigate the X-ray properties of the Parkes sample of flat-spectrum 
radio sources using data from the {\em ROSAT} All--Sky Survey
and archival pointed PSPC observations. In total, 163 of the 323 sources 
are detected. For the remaining 160 sources 2$\sigma$ upper limits to 
the X-ray flux are derived. We present power-law photon indices in the 
0.1--2.4 keV energy band for 115 sources, which were either determined 
with a hardness ratio technique or from direct fits to pointed PSPC 
data if a sufficient number of photons is available. The average photon 
index is $\langle\Gamma\rangle = 1.95^{+0.13}_{-0.12}$ for flat-spectrum 
radio-loud quasars, $\langle\Gamma\rangle = 1.70^{+0.23}_{-0.24}$ for 
galaxies, and $\langle\Gamma\rangle = 2.40^{+0.12}_{-0.31}$ for BL Lac 
objects. The soft X-ray photon index is correlated with redshift and with 
radio spectral index in the sense that sources at high redshift and/or with 
flat (or inverted) radio spectra have flatter X-ray spectra on average. The 
results are in accord with orientation dependent unification schemes
for radio-loud AGN.

Webster et al. \shortcite{webster} discovered many sources with unusually red 
optical continua among the quasars of this sample and interpreted this result 
in terms of extinction by dust. Although the X-ray spectra in general do not 
show excess absorption, we find that low-redshift optically red quasars have
significantly lower soft X-ray luminosities on average than objects with blue
optical continua. The difference disappears for higher redshifts, as is
expected for intrinsic absorption by cold gas associated with the dust. 
In addition, the scatter in $\log (f_{\rm x}/f_{\rm o})$ is consistent with
the observed optical extinction, contrary to previous claims based on
optically or X-ray selected samples. 

Although alternative explanations for the red optical continua cannot be 
excluded with the present X-ray data, we note that the observed X-ray 
properties are consistent with the idea that dust plays an important role 
in some of the radio-loud quasars with red optical continua.  

\end{abstract}


\section{Introduction}
Between August 1990 and February 1991 {\em ROSAT} \cite{truemper}
performed a survey of the whole sky in the soft X-ray band between 
0.1--2.4 keV \cite{voges} with the Position Sensitive Proportional 
Counter \cite{pfeffermann}. This survey yielded about 60\,000 
X-ray sources 
with integrated fluxes greater than a few times 10$^{\rm -13}$ \ergs.
One of the outstanding achievements of this survey is that it offers
the possibility to investigate the X-ray properties of large samples of 
astrophysical objects in an unbiased way.

Many well-defined samples of Active Galactic Nuclei (AGN), selected from 
different wavebands, have meanwhile been studied in the X-ray regime. 
Among others, the Large Bright Quasar Sample \cite{green},
the Molonglo quasar sample \cite{baker}, the 3CRR catalog 
\cite{prieto} and the Wall \& Peacock 2-Jy sample \cite{siebert}. 
In addition, the {\em ROSAT} All--Sky Survey (RASS) has been
correlated with large radio catalogs like the Molonglo Reference Catalog
(Large et al. 1981; Brinkmann, Siebert \& Boller 1994) and the Green Bank
5-GHz survey (Condon, Broderick \& Seielstad 1989; Brinkmann et al. 1995, 
1997b) as well as
the V\'eron quasar catalog (V\'eron-Cetty \& V\'eron 1993; Brinkmann, 
Yuan \& Siebert 1997a; Yuan, Brinkmann \& Siebert 1998). For the first time 
it is possible
to compare the X-ray properties of large samples of AGN selected 
at different wavelengths. These studies make an important contribution to
our understanding of the phenomenological differences between the various 
types of AGN and the underlying physical processes.

In this paper we investigate the X-ray properties of the Parkes sample
of flat-spectrum radio sources \cite{drinkwater}. The choice of this sample
was particularly motivated by the recent discovery of a large number of
sources with extremely red optical to infrared continua in this 
sample \cite{webster}. It was suggested that this reddening is due to dust 
in the line-of-sight and it therefore seemed natural to look for the
imprints of this material on the soft X-ray emission by absorption at very 
low energies.

Moreover, it is interesting to compare the X-ray properties of flat-spectrum 
quasars with those of steep-spectrum quasars, like, for example, the Molonglo 
quasar sample \cite{baker} or the steep-spectrum quasars of Brinkmann 
et al. \shortcite{bys}. Since orientation based unification schemes for 
radio-loud quasars predict that both classes belong to the same parent 
population, but with different orientations to the line-of-sight, it is 
important to see how the X-ray properties fit into this scenario.
 
The outline of the paper is as follows. In Section 2 we introduce the 
Parkes catalog of flat-spectrum radio sources and describe the analysis
of the X-ray data from the RASS and pointed PSPC observations in Section 3.
Section 4 deals with the results of our analysis in terms of detection rates,
variability, X-ray spectra, and luminosity correlations. In Section 5 we
focus on the X-ray evidence for dust reddened quasars. A discussion of our
results and the conclusions are presented in Section 6 and 7, respectively. 

\section{The catalog}

The original Parkes catalog contains about 10\,000 radio sources and was 
the result of a survey of the southern sky with the Parkes radio telescope 
at 2.7\,GHz between 1968 and 1979 (Bolton, Savage \& Wright 1979; Wright 
\& Otrupcek 1990). To compile a large and unbiased sample of radio-selected 
quasars, sources were selected from the original survey papers according 
to the following criteria \cite{drinkwater}:
$S_{\rm 2.7\,GHz} >$ 0.5 Jy,
radio spectral index $\alpha^{\rm 5}_{\rm 2.7} > -$0.5 (where 
$S_{\nu} \propto \nu^{\alpha}$),
Galactic latitude $|b| >$ 20$\degr$,
$-45\degr < \delta_{\rm 1950.0} < +$10$\degr$,
2.7\,GHz ($S_{\rm 2.7\,GHz}$) and 5\,GHz flux density ($S_{\rm 5\,GHz}$)
available in the papers.

By selecting flat-spectrum radio sources the resulting sample is biased
towards core-dominated quasars, because radio galaxies and lobe-dominated 
quasars generally have steeper radio spectra. The final complete list contains
323 sources. Special care has been taken in the optical identification of
the objects and in the determination of their redshifts. Using 
optical databases and new observations, 321 sources could be identified
and for 277 of them redshifts are available. 

Drinkwater et al. \shortcite{drinkwater} used digitized optical plates to 
automatically classify the morphologies as 237 'stellar' (unresolved), 
35 'galaxy' (resolved), 11 'merged' (two or more images too close together 
on the optical plate to be separated) and 38 'faint' sources (too faint to
classify). They used CCD images to deconvolve and identify the 11
'merged' objects. All of these were 'stellar' except PKS~1555$-$140
which was a 'galaxy'. For the purposes of this paper we define an
additional category of 14 'BL Lac objects' being the 12 'stellar', one
'galaxy' and one 'merged' objects that are {\em bona fide} BL Lac
objects according to Padovani \& Giommi \shortcite{padovani}\footnote{In 
detail, the 14 BL Lac objects are: 0048$-$097, 0118$-$272, 0138$-$097, 
0301$-$243, 0422$+$004, 0426$-$380, 0537$-$441, 0823$+$033, 0829$+$046, 
1144$-$379, 1514$-$241, 1519$-$273, 2131$-$021, 2240$-$260}. These objects 
will be handled separately in the following subsections. We divide the 
remaining sources as follows: 234 stellar sources ('quasars'), 35 'galaxies' 
and 38 'faint' sources. Note that the quasar/galaxy distinction is only 
based on morphology so that a quasar with an associated nebulosity like 
PKS~0445+097 is classified as a 'galaxy'. For a discussion of the 
completeness of the sample and a detailed presentation of the radio and 
optical data see Drinkwater et al. \shortcite{drinkwater}.

\section{Data analysis}

\subsection{{\em ROSAT} All--Sky Survey data}

For each of the sample sources a 1$^{\circ}\times$1$^{\circ}$ field
centered on the radio position was extracted from the {\em ROSAT} All--Sky
Survey and analysed using a procedure based on standard routines 
within the {\sevensize EXSAS} environment \cite{zimmermann}. This 
procedure uses a maximum-likelihood source detection algorithm which 
returns the likelihood of existence for a X-ray source at the specified 
radio position, the number of source photons within 5 times the FWHM of 
the PSPC's point spread function and the error in the number of source 
photons. For the {\em ROSAT} All--Sky Survey the FWHM of the PSPC point 
spread function is estimated to be $\sim 60$ arcsec \cite{zimmermann}. 

The choice of the background has a significant influence on the results
of the source detection procedure, in particular for weak sources.
We estimated the local background by taking the average of two source
free boxes, each 10 arcmin by 10 arcmin in size, which are offset
by 15 arcmin from the radio position along the scanning direction of
the satellite during the All--Sky Survey observations. In this way it is
ensured that the background regions have an exposure similar to the source
region. 

Since it is known that an AGN is present at the position of the radio 
source, we considered a radio source to be detected in X-rays if the 
likelihood of existence is greater than 5.91, which corresponds 
to 3$\sigma$. If no X-ray source is detected above the specified significance 
level, we determined the 2$\sigma$ upper limit on the number of X-ray 
photons. To calculate the corresponding count rates we used the 
vignetting-corrected RASS exposure averaged over a circle with radius 
5 arcmin centered on the radio position.

The unabsorbed fluxes are calculated using the standard {\it Energy to Counts
Conversion Factors} (ECF) assuming a simple power-law model modified by 
Galactic absorption. We used the photon indices $\Gamma$\footnote{$\Gamma$ 
is the power-law slope of the X-ray photon spectrum defined as 
N(E)$\propto$E$^{-\Gamma}$. It is connected to the energy index via $\alpha 
= \Gamma - 1$.} listed in Tables 1 and 2, if the uncertainty is smaller 
than 0.5. For the other sources we applied $\Gamma$ = 2.0 for quasars and 
$\Gamma$ = 1.7 for galaxies. These values represent the average spectral 
indices of the respective object classes derived in Section 4.4. 
 
\subsection{Pointed {\em ROSAT} observations}

In addition to the RASS data we searched the {\em ROSAT} source catalog
(ROSAT-SRC; Voges et al. 1994), which was generated from pointed
PSPC observations. We found 49 sources which were in the field of view of 
an observation either as the main target or serendipitously. Of these, 12
were detected in the pointed observation only. Fluxes and luminosities were
calculated based on the count rates from the catalog and assuming the same 
spectral parameters as for the RASS data. In the analysis we used the fluxes
from the pointed observations only when no Survey detection is available.
 
\subsection{Spectral analysis}

In general, simple power-law models for the X-ray spectra modified by 
neutral gas absorption were considered in the analysis. The amount of 
absorption is parametrized by the column density of neutral hydrogen 
in the line of sight to the source (the $N_{\rm H}$ value), assuming 
standard element abundances and the energy dependent absorption cross 
sections of Morrison \& McCammon \shortcite{morrison}. The Galactic 
$N_{\rm H}$ values are determined from radio measurements \cite{dickey}. 
For each source two photon indices were determined, one by fixing the 
absorption to the Galactic value, the other one by leaving the amount 
of absorption as a free parameter.

For weak sources (mostly from the All--Sky Survey) a technique based on 
hardness ratios was applied to estimate the spectral parameters. The two 
standard {\em ROSAT} hardness ratios are defined as $HR_{\rm 1} = 
(B - A)/(B + A)$ and $HR_{\rm 2} = (D - C)/(D + C)$, where {\em A}, {\em B}, 
{\em C} and {\em D} are the number of photons in the pulse height channel 
intervals 11--41, 52--201, 52--90 and 91--201, respectively. Every 
combination of photon index $\Gamma$ and absorption $N_{\rm H}$ leads to unique
values for the two hardness ratios when folded with the instrument response.
The inversion of this procedure then allows to determine the spectral 
parameters from the measured hardness ratios. However, this method only gives 
a rough estimate of the spectral parameters, in particular for the two 
parameter fits. A detailed discussion of the method is presented in Schartel
\shortcite{schartel}. In total we obtained spectral information for 104
sources in this way.

Whenever a pointed PSPC observation was available for a source from the
public archive and if a sufficient number of source photons (\ma 50) was 
accumulated in this pointing, we determined the spectral parameters by 
explicitly fitting a power-law model to the data. X-ray spectra of 42 sources 
were determined in this way with increased accuracy. The results are
presented in Table 2. A few sources were observed more than once with the 
PSPC (e.g. 3C 279). In these cases the observation with the longest exposure 
was chosen for the spectral analysis. If the number of photons was too low 
for spectral fitting we again used the hardness ratios (this time derived 
from the {\it pointed} observation) to determine the spectral parameters.

In total, spectral information could be obtained for 115 sources of the 
sample. The X-ray spectra are discussed in Section 4.4.

\begin{table*}   
\small
\tabcolsep1.2ex
\caption{X-ray properties of the Parkes sample of flat-spectrum radio sources}
\begin{tabular}{rcrrrcccrcrcl} 
\noalign{\smallskip} \hline \noalign{\smallskip}  
\multicolumn{1}{c}{Object}       & \multicolumn{1}{c}{Type}            & 
\multicolumn{1}{c}{m$_{B}$}      & \multicolumn{1}{c}{z}               & 
\multicolumn{1}{c}{f$_{2.7GHz}$} &                                     & 
\multicolumn{1}{c}{cts/s}        & \multicolumn{1}{c}{$F_x$}           & 
\multicolumn{1}{c}{N$_{H,gal}$}  & \multicolumn{1}{c}{$\Gamma_{gal}$}  & 
\multicolumn{1}{c}{N$_{H,free}$} & \multicolumn{1}{c}{$\Gamma_{free}$} & \\
\multicolumn{1}{c}{(1)}          & \multicolumn{1}{c}{(2)}             &  
\multicolumn{1}{c}{(3)}          & \multicolumn{1}{c}{(4)}             &  
\multicolumn{1}{c}{(5)}          & \multicolumn{1}{c}{(6)}             & 
\multicolumn{1}{c}{(7)}          & \multicolumn{1}{c}{(8)}             & 
\multicolumn{1}{c}{(9)}          & \multicolumn{1}{c}{(10)}            & 
\multicolumn{1}{c}{(11)}         & \multicolumn{1}{c}{(12)}   & \multicolumn{1}{l}{(13)}\\ 
\noalign{\smallskip} \hline \noalign{\smallskip}
 0003$-$066  & s & 18.5 & 0.347 &  1.460 & S  & $ $0.070$\pm$ 0.016 & $ $  1.300$\pm$ 0.305 &  3.14 & 1.58$_{-0.92}^{+0.48}$ &  ...                      &  ...                   &  \\
 0005$-$239  & s & 16.6 & 1.407 &  0.580 & S  & $<$0.018            & $<$  0.278            &  2.29 &  ...                   &  ...                      &  ...                   &  \\
 0005$-$262  & s &  0.0 & 0.000 &  0.580 & S  & $<$0.011            & $<$  0.145            &  1.74 &  ...                   &  ...                      &  ...                   &  \\
 0008$-$264  & s & 19.5 & 1.096 &  0.670 & S  & $<$0.013            & $<$  0.196            &  2.19 &  ...                   &  ...                      &  ...                   &  \\
 0013$-$005  & s & 19.4 & 1.574 &  0.890 & S  & $<$0.019            & $<$  0.333            &  2.91 &  ...                   &  ...                      &  ...                   &  \\
\end{tabular}
\end{table*}
\setcounter{table}{0}
\begin{table*}   
\small
\tabcolsep1.2ex
\caption{continued}
\begin{tabular}{rcrrrcccrcrcl}
\noalign{\smallskip} \hline \noalign{\smallskip}  
\multicolumn{1}{c}{Object}       & \multicolumn{1}{c}{Type}            & 
\multicolumn{1}{c}{m$_{B}$}      & \multicolumn{1}{c}{z}               & 
\multicolumn{1}{c}{f$_{2.7GHz}$} &                                     & 
\multicolumn{1}{c}{cts/s}        & \multicolumn{1}{c}{$F_x$}           & 
\multicolumn{1}{c}{N$_{H,gal}$}  & \multicolumn{1}{c}{$\Gamma_{gal}$}  & 
\multicolumn{1}{c}{N$_{H,free}$} & \multicolumn{1}{c}{$\Gamma_{free}$} & \\
\multicolumn{1}{c}{(1)}          & \multicolumn{1}{c}{(2)}             &  
\multicolumn{1}{c}{(3)}          & \multicolumn{1}{c}{(4)}             &  
\multicolumn{1}{c}{(5)}          & \multicolumn{1}{c}{(6)}             & 
\multicolumn{1}{c}{(7)}          & \multicolumn{1}{c}{(8)}             & 
\multicolumn{1}{c}{(9)}          & \multicolumn{1}{c}{(10)}            & 
\multicolumn{1}{c}{(11)}         & \multicolumn{1}{c}{(12)}            & 
\multicolumn{1}{l}{(13)} \\ 
\noalign{\smallskip} \hline \noalign{\smallskip}  
 0338$-$214  & s & 16.0 & 0.048 &  0.820 & S  & $ $0.017$\pm$ 0.008 & $ $  0.297$\pm$ 0.141 &  2.79 &  ...                   &  ...                      &  ...                   &  \\
 0346$-$163  & s & 17.4 & 0.000 &  0.540 & S  & $<$0.007            & $<$  0.153            &  4.05 &  ...                   &  ...                      &  ...                   &  \\
 0346$-$279  & s & 20.5 & 0.987 &  1.100 & S  & $ $0.132$\pm$ 0.025 & $ $  1.290$\pm$ 0.241 &  0.96 & 2.55$_{-0.28}^{+0.35}$ &  1.91$_{- 1.80}^{+ 8.46}$ & 3.00$_{-1.28}^{+3.00}$ &  \\
 0348$+$049  & f &  0.0 & 0.000 &  0.540 & S  & $<$0.009            & $<$  0.293            & 13.29 &  ...                   &  ...                      &  ...                   &  \\
 0348$-$120  & s & 17.9 & 1.520 &  0.500 & S  & $<$0.009            & $<$  0.172            &  3.45 &  ...                   &  ...                      &  ...                   &  \\
\end{tabular}
\end{table*}
\setcounter{table}{0}
\begin{table*}   
\small
\tabcolsep1.2ex
\caption{continued}
\begin{tabular}{rcrrrcccrcrcl}
\noalign{\smallskip} \hline \noalign{\smallskip}  
\multicolumn{1}{c}{Object}       & \multicolumn{1}{c}{Type}            & 
\multicolumn{1}{c}{m$_{B}$}      & \multicolumn{1}{c}{z}               & 
\multicolumn{1}{c}{f$_{2.7GHz}$} &                                     & 
\multicolumn{1}{c}{cts/s}        & \multicolumn{1}{c}{$F_x$}           & 
\multicolumn{1}{c}{N$_{H,gal}$}  & \multicolumn{1}{c}{$\Gamma_{gal}$}  & 
\multicolumn{1}{c}{N$_{H,free}$} & \multicolumn{1}{c}{$\Gamma_{free}$} & \\
\multicolumn{1}{c}{(1)}          & \multicolumn{1}{c}{(2)}             &  
\multicolumn{1}{c}{(3)}          & \multicolumn{1}{c}{(4)}             &  
\multicolumn{1}{c}{(5)}          & \multicolumn{1}{c}{(6)}             & 
\multicolumn{1}{c}{(7)}          & \multicolumn{1}{c}{(8)}             & 
\multicolumn{1}{c}{(9)}          & \multicolumn{1}{c}{(10)}            & 
\multicolumn{1}{c}{(11)}         & \multicolumn{1}{c}{(12)}            & 
\multicolumn{1}{l}{(13)} \\ 
\noalign{\smallskip} \hline \noalign{\smallskip}  
 0925$-$203  & s & 16.4 & 0.348 &  0.810 & S  & $ $0.167$\pm$ 0.021 & $ $  3.700$\pm$ 0.463 &  3.71 & 2.17$_{-0.20}^{+0.18}$ &  3.62$_{- 2.78}^{+ 3.39}$ & 2.14$_{-1.01}^{+1.02}$ &  \\
 1004$-$018  & s & 20.3 & 1.212 &  0.560 & S  & $<$0.013            & $<$  0.260            &  3.70 &  ...                   &  ...                      &  ...                   &  \\
 1008$-$017  & s & 19.6 & 0.406 &  0.800 & S  & $<$0.026            & $<$  0.527            &  3.70 &  ...                   &  ...                      &  ...                   &  \\
 1016$-$311  & s & 17.6 & 0.794 &  0.620 & S  & $<$0.016            & $<$  0.380            &  5.07 &  ...                   &  ...                      &  ...                   &  \\
 1020$-$103  & s & 15.1 & 0.197 &  0.640 & S  & $ $0.182$\pm$ 0.023 & $ $  3.830$\pm$ 0.479 &  4.84 & 1.74$_{-0.30}^{+0.24}$ &  3.48$_{- 3.09}^{+ 4.14}$ & 1.40$_{-1.04}^{+1.00}$ &  \\
\end{tabular}
\end{table*}
\setcounter{table}{0}
\begin{table*}   
\small
\tabcolsep1.2ex
\caption{continued}
\begin{tabular}{rcrrrcccrcrcl}
\noalign{\smallskip} \hline \noalign{\smallskip}  
\multicolumn{1}{c}{Object}       & \multicolumn{1}{c}{Type}            & 
\multicolumn{1}{c}{m$_{B}$}      & \multicolumn{1}{c}{z}               & 
\multicolumn{1}{c}{f$_{2.7GHz}$} &                                     & 
\multicolumn{1}{c}{cts/s}        & \multicolumn{1}{c}{$F_x$}           & 
\multicolumn{1}{c}{N$_{H,gal}$}  & \multicolumn{1}{c}{$\Gamma_{gal}$}  & 
\multicolumn{1}{c}{N$_{H,free}$} & \multicolumn{1}{c}{$\Gamma_{free}$} & \\
\multicolumn{1}{c}{(1)}          & \multicolumn{1}{c}{(2)}             &  
\multicolumn{1}{c}{(3)}          & \multicolumn{1}{c}{(4)}             &  
\multicolumn{1}{c}{(5)}          & \multicolumn{1}{c}{(6)}             & 
\multicolumn{1}{c}{(7)}          & \multicolumn{1}{c}{(8)}             & 
\multicolumn{1}{c}{(9)}          & \multicolumn{1}{c}{(10)}            & 
\multicolumn{1}{c}{(11)}         & \multicolumn{1}{c}{(12)}            & 
\multicolumn{1}{l}{(13)} \\ 
\noalign{\smallskip} \hline \noalign{\smallskip}
 1302$-$102  & s & 15.7 & 0.286 &  0.890 & SP & $ $0.323$\pm$ 0.036 & $ $  7.350$\pm$ 0.816 &  3.38 & 2.62$_{-0.18}^{+0.17}$ &  2.85$_{- 2.59}^{+ 3.41}$ & 2.42$_{-1.10}^{+1.15}$ & *\\
 1313$-$333  & s & 16.8 & 1.210 &  1.000 & S  & $<$0.020            & $<$  0.453            &  4.84 &  ...                   &  ...                      &  ...                   &  \\
 1317$+$019  & s & 20.8 & 1.232 &  0.550 & UP & $ $0.021$\pm$ 0.003 & $ $  0.314$\pm$ 0.039 &  1.95 & 1.64$_{-0.22}^{+0.20}$ &  3.05$_{- 2.95}^{+ 3.90}$ & 2.01$_{-1.17}^{+1.18}$ &  \\
 1318$-$263  & s & 20.4 & 2.027 &  0.650 & S  & $<$0.015            & $<$  0.381            &  6.36 &  ...                   &  ...                      &  ...                   &  \\
 1327$-$311  & s & 18.5 & 1.335 &  0.520 & S  & $<$0.013            & $<$  0.255            &  3.63 &  ...                   &  ...                      &  ...                   &  \\
\end{tabular}
\end{table*}
\setcounter{table}{0}
\begin{table*}   
\small
\tabcolsep1.2ex
\caption{continued}
\begin{tabular}{rcrrrcccrcrcl}
\noalign{\smallskip} \hline \noalign{\smallskip}  
\multicolumn{1}{c}{Object}       & \multicolumn{1}{c}{Type}            & 
\multicolumn{1}{c}{m$_{B}$}      & \multicolumn{1}{c}{z}               & 
\multicolumn{1}{c}{f$_{2.7GHz}$} &                                     & 
\multicolumn{1}{c}{cts/s}        & \multicolumn{1}{c}{$F_x$}           & 
\multicolumn{1}{c}{N$_{H,gal}$}  & \multicolumn{1}{c}{$\Gamma_{gal}$}  & 
\multicolumn{1}{c}{N$_{H,free}$} & \multicolumn{1}{c}{$\Gamma_{free}$} & \\
\multicolumn{1}{c}{(1)}          & \multicolumn{1}{c}{(2)}             &  
\multicolumn{1}{c}{(3)}          & \multicolumn{1}{c}{(4)}             &  
\multicolumn{1}{c}{(5)}          & \multicolumn{1}{c}{(6)}             & 
\multicolumn{1}{c}{(7)}          & \multicolumn{1}{c}{(8)}             & 
\multicolumn{1}{c}{(9)}          & \multicolumn{1}{c}{(10)}            & 
\multicolumn{1}{c}{(11)}         & \multicolumn{1}{c}{(12)}            & 
\multicolumn{1}{l}{(13)} \\ 
\noalign{\smallskip} \hline \noalign{\smallskip}
 1614$+$051  & s & 21.1 & 3.217 &  0.670 & S  & $ $0.023$\pm$ 0.009 & $ $  0.520$\pm$ 0.212 &  4.90 &  ...                   &  ...                      &  ...                   &  \\
 1615$+$029  & s & 18.2 & 1.339 &  0.740 & S  & $<$0.018            & $<$  0.431            &  5.55 &  ...                   &  ...                      &  ...                   &  \\
 1616$+$063  & s & 19.6 & 2.086 &  0.930 & S  & $<$0.017            & $<$  0.383            &  4.63 &  ...                   &  ...                      &  ...                   &  \\
 1635$-$035  & s & 21.8 & 2.856 &  0.510 & S  & $<$0.008            & $<$  0.259            & 10.79 &  ...                   &  ...                      &  ...                   &  \\
 1648$+$015  & f & 22.7 & 0.000 &  0.720 & S  & $<$0.008            & $<$  0.192            &  6.14 &  ...                   &  ...                      &  ...                   &  \\
\end{tabular}
\end{table*}
\setcounter{table}{0}
\begin{table*}   
\small
\tabcolsep1.2ex
\caption{continued}
\begin{tabular}{rcrrrcccrcrcl}
\noalign{\smallskip} \hline \noalign{\smallskip}  
\multicolumn{1}{c}{Object}       & \multicolumn{1}{c}{Type}            & 
\multicolumn{1}{c}{m$_{B}$}      & \multicolumn{1}{c}{z}               & 
\multicolumn{1}{c}{f$_{2.7GHz}$} &                                     & 
\multicolumn{1}{c}{cts/s}        & \multicolumn{1}{c}{$F_x$}           & 
\multicolumn{1}{c}{N$_{H,gal}$}  & \multicolumn{1}{c}{$\Gamma_{gal}$}  & 
\multicolumn{1}{c}{N$_{H,free}$} & \multicolumn{1}{c}{$\Gamma_{free}$} & \\
\multicolumn{1}{c}{(1)}          & \multicolumn{1}{c}{(2)}             &  
\multicolumn{1}{c}{(3)}          & \multicolumn{1}{c}{(4)}             &  
\multicolumn{1}{c}{(5)}          & \multicolumn{1}{c}{(6)}             & 
\multicolumn{1}{c}{(7)}          & \multicolumn{1}{c}{(8)}             & 
\multicolumn{1}{c}{(9)}          & \multicolumn{1}{c}{(10)}            & 
\multicolumn{1}{c}{(11)}         & \multicolumn{1}{c}{(12)}            & 
\multicolumn{1}{l}{(13)} \\ 
\noalign{\smallskip} \hline \noalign{\smallskip}
 2227$-$088  & s & 18.0 & 1.561 &  1.490 & S  & $<$0.014            & $<$  0.314            &  4.51 &  ...                   &  ...                      &  ...                   &  \\
 2227$-$399  & s & 17.4 & 0.323 &  1.020 & SP & $ $0.509$\pm$ 0.043 & $ $  5.750$\pm$ 0.481 &  1.20 & 1.98$_{-0.12}^{+0.11}$ &  2.01$_{- 1.91}^{+ 2.58}$ & 2.31$_{-0.92}^{+0.96}$ & *\\
 2229$-$172  & f & 21.3 & 1.780 &  0.520 & S  & $<$0.014            & $<$  0.272            &  3.26 &  ...                   &  ...                      &  ...                   &  \\
 2233$-$148  & s & 20.9 & 0.609 &  0.500 & S  & $ $0.039$\pm$ 0.016 & $ $  0.866$\pm$ 0.354 &  4.46 &  ...                   &  ...                      &  ...                   &  \\
 2239$+$096  & s & 19.7 & 1.707 &  0.650 & S  & $ $0.028$\pm$ 0.010 & $ $  0.699$\pm$ 0.260 &  5.89 & 2.87$_{-1.00}^{+0.71}$ & 11.73$_{-11.56}^{+21.40}$ & 4.43$_{-4.21}^{+1.56}$ &  \\
\hline
\end{tabular}

\medskip

\begin{minipage}{17cm}
Column (1) Object name. Objects in brackets do not belong to the complete 
sample. Column (2) Identification from plate according to Drinkwater et al. 
(1997) and Peterson (priv. com.). 's' = stellar, 'g' = galaxy, 'f' = faint, 
'X' = crowded field, 'D' = double galaxy, 'B' = BL Lac candidate,
'Q' = quasar, 'E' = elliptical galaxy. Column (3) Blue magnitude ($B_{\rm J}$)
according to Drinkwater et al. (1997). Column (4) Redshift. Column (5) Radio flux density 
at 2.7 GHz in Jy. Column (6) Reference to the X-ray detection. 'S' = Survey, 
'P' = pointed observation. Column (7) X-ray count rate in the ROSAT energy 
band (0.1--2.4 keV). In general the Survey count rate is given, except for the 
sources that are detected in pointed observations only. Upper limits are 90 per cent 
confidence. Column (8) Total 0.1--2.4 keV X-ray flux in units of 
10$^{\rm -12}$ erg cm$^{-2}$ s$^{-1}$. Calculation was done assuming Galactic 
absorption only and the average X-ray photon index for radio-loud quasars 
($\Gamma = 2.0$). The individual photon index for a source was used in the 
calculation whenever it was available and well determined (i.e., 
$\Delta\Gamma < 0.5$). Column (9) Galactic $N_{\rm H}$ value in 
10$^{\rm 20}$ cm$^{\rm -2}$ (Dickey \& Lockman 1990). Column (10) X-ray 
photon index $\Gamma_{\rm gal}$. Errors are 1$\sigma$.
For sources detected in pointed observations it was determined by an explicit 
fit of a power-law model to the data. For sources detected in the All--Sky 
Survey only it was computed using a hardness-ratio technique (Schartel 1996).
In both cases the Galactic $N_{\rm H}$ value was assumed. Columns (11),(12) 
Results of the two parameter spectral fits. The fitted absorption ($N_{\rm 
H,free}$) and the respective power-law index ($\Gamma_{\rm free}$) are given. 
Column (13) An asterisk denotes the sources, which have spectral parameters available
also from explicit fits to the data. The results of the fits are presented in 
Table 2.
\end{minipage}

\end{table*}                          

\begin{figure}
\hspace*{-0.8cm}
\psfig{figure=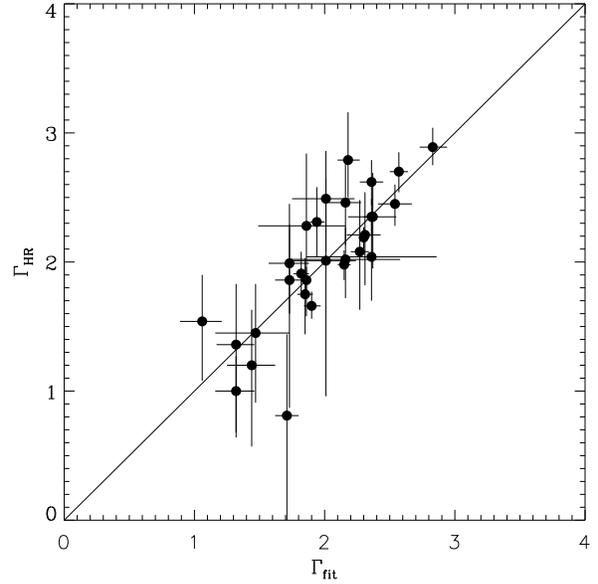,width=9cm}
\caption[]{\label{hrfit} Comparison of the photon indices derived from the
hardness ratios with those from spectral fits to pointed observations for
the 30 objects, where both, RASS and pointed data, were available. Only the
fits for fixed galactic $N_{\rm H}$ are shown.}
\end{figure}

For 30 sources we have spectral information from both the RASS and
pointed observations. In order to validate the hardness ratio method we
compare in Fig. \ref{hrfit} the photon indices derived from the hardness
ratios with those determined from spectral fits. In general, the
two independently derived photon indices agree very well. 
For all sources the results agree within their respective 2\,$\sigma$ errors. 

\subsection{The Tables}
The X-ray properties of the Parkes sample of flat-spectrum radio sources
are summarized in Table 1. For a detailed description of its contents we refer
to the table notes. In columns (10) to (12) we give the spectral parameters 
as derived from the hardness ratios only,
either from the RASS or (in some cases) from pointed observations. In 
column (13) we indicate if a detailed fit is available. The parameters
derived from spectral fitting are given in Table 2.

\setcounter{table}{1}
\begin{table}
\small
\tabcolsep1ex                           
\caption{\label{fits} Results from spectral fitting.}                     
\begin{tabular}{lclccl}
\noalign{\smallskip} \hline \noalign{\smallskip}
\multicolumn{1}{c}{Name} & \multicolumn{1}{c}{$\Gamma_{\rm gal}$} & 
\multicolumn{1}{c}{$\chi^2_{red}$} & \multicolumn{1}{c}{$\Gamma_{\rm free}$} 
& \multicolumn{1}{c}{N$_{\rm H,free}$} & \multicolumn{1}{c}{$\chi^2_{red}$} \\
\multicolumn{1}{c}{(1)} & \multicolumn{1}{c}{(2)} & \multicolumn{1}{c}{(3)} 
&\multicolumn{1}{c}{(4)} & \multicolumn{1}{c}{(5)} & \multicolumn{1}{c}{(6)}\\
\noalign{\smallskip} \hline \noalign{\smallskip}
 0048$-$097  & 2.57$_{-0.07}^{+0.07}$ & 0.82(92) & 2.79$_{-0.24}^{+0.24}$ &  4.20$_{- 0.68}^{+ 0.72}$ & 0.77(91)\\
 0118$-$272  & 2.39$_{-0.14}^{+0.14}$ & 0.93(15) & 2.84$_{-0.50}^{+0.56}$ &  2.79$_{- 1.25}^{+ 1.56}$ & 0.81(14)\\
 0122$-$003  & 2.16$_{-0.13}^{+0.12}$ & 0.92(24) & 1.72$_{-0.28}^{+0.30}$ &  1.96$_{- 0.67}^{+ 0.86}$ & 0.67(23)\\
 0135$-$247  & 2.18$_{-0.08}^{+0.09}$ & 1.20(28) & 1.82$_{-0.20}^{+0.22}$ &  0.59$_{- 0.35}^{+ 0.42}$ & 0.94(27)\\
 0137$+$012  & 1.94$_{-0.05}^{+0.06}$ & 1.08(20) & 1.99$_{-0.16}^{+0.16}$ &  3.21$_{- 0.49}^{+ 0.52}$ & 1.12(19)\\
 0232$-$042  & 2.27$_{-0.07}^{+0.07}$ & 1.15(32) & 1.96$_{-0.19}^{+0.20}$ &  1.74$_{- 0.47}^{+ 0.52}$ & 0.96(31)\\
 0238$-$084  & 1.53$_{-0.17}^{+0.15}$ & 0.88(15) & 1.58$_{-0.37}^{+0.36}$ &  3.30$_{- 1.35}^{+ 1.62}$ & 0.98(14)\\
 0301$-$243  & 2.83$_{-0.10}^{+0.11}$ & 0.91(23) & 2.77$_{-0.27}^{+0.31}$ &  1.62$_{- 0.53}^{+ 0.62}$ & 0.95(22)\\
 0316$-$444  & 1.86$_{-0.02}^{+0.02}$ & 1.99(80) & 1.56$_{-0.07}^{+0.07}$ &  1.56$_{- 0.20}^{+ 0.20}$ & 1.14(79)\\
 0403$-$132  & 1.71$_{-0.09}^{+0.09}$ & 0.95(22) & 1.75$_{-0.23}^{+0.23}$ &  3.80$_{- 0.90}^{+ 0.90}$ & 0.99(21)\\
 0405$-$123  & 2.36$_{-0.06}^{+0.05}$ & 0.91(19) & 2.23$_{-0.16}^{+0.17}$ &  3.46$_{- 0.50}^{+ 0.53}$ & 0.87(18)\\
 0426$-$380  & 1.95$_{-0.27}^{+0.27}$ & 2.93(6)  & 3.20$_{-1.16}^{+1.35}$ &  5.28$_{- 3.02}^{+ 3.82}$ & 2.36(5) \\
 0430$+$052  & 1.85$_{-0.06}^{+0.06}$ & 1.85(51) & 2.52$_{-0.24}^{+0.30}$ & 19.66$_{- 3.50}^{+ 4.68}$ & 0.95(50)\\
 0438$-$436  & 0.70$_{-0.10}^{+0.10}$ & 2.32(19) & 1.96$_{-0.51}^{+1.48}$ & 11.00$_{- 5.35}^{+26.43}$ & 0.62(18)\\
 0500$+$019  & 0.71$_{-0.63}^{+0.56}$ & 0.93(8)  & 2.22$_{-2.04}^{+2.91}$ & 39.52$_{-35.46}^{+60.47}$ & 0.91(7) \\
 0537$-$441  & 2.54$_{-0.13}^{+0.13}$ & 1.56(35) & 2.04$_{-0.32}^{+0.32}$ &  2.54$_{- 0.85}^{+ 0.99}$ & 1.30(34)\\
 0537$-$286  & 1.32$_{-0.16}^{+0.14}$ & 1.11(15) & 1.59$_{-0.35}^{+0.36}$ &  3.11$_{- 1.23}^{+ 1.48}$ & 1.06(14)\\
 1034$-$293  & 2.01$_{-0.26}^{+0.22}$ & 1.76(16) & 1.27$_{-0.38}^{+0.39}$ &  1.27$_{- 0.93}^{+ 1.17}$ & 1.08(15)\\
 1055$+$018  & 2.37$_{-0.19}^{+0.18}$ & 1.42(11) & 1.64$_{-0.43}^{+0.47}$ &  1.62$_{- 1.08}^{+ 1.36}$ & 0.85(10)\\
 1136$-$135  & 2.36$_{-0.50}^{+0.50}$ & 1.00(10) & 2.24$_{-1.58}^{+1.58}$ &  3.20$_{- 3.20}^{+ 5.70}$ & 1.11(9) \\
 1145$-$071  & 1.80$_{-0.82}^{+0.55}$ & 0.28(4)  & 1.52$_{-1.08}^{+1.06}$ &  2.30$_{- 2.30}^{+ 8.99}$ & 0.30(3) \\
 1144$-$379  & 2.54$_{-0.19}^{+0.17}$ & 0.81(34) & 2.67$_{-0.37}^{+0.39}$ &  9.48$_{- 1.88}^{+ 2.50}$ & 0.81(33)\\
 1226$+$023  & 2.30$_{-0.02}^{+0.01}$ & 1.31(56) & 2.16$_{-0.06}^{+0.05}$ &  1.45$_{- 0.12}^{+ 0.13}$ & 1.01(55)\\
 1253$-$055  & 1.90$_{-0.06}^{+0.07}$ & 0.60(21) & 1.89$_{-0.19}^{+0.20}$ &  2.23$_{- 0.57}^{+ 0.62}$ & 0.63(20)\\
 1302$-$102  & 2.36$_{-0.09}^{+0.09}$ & 0.71(29) & 2.12$_{-0.24}^{+0.26}$ &  2.63$_{- 0.70}^{+ 0.77}$ & 0.65(28)\\
 1330$+$022  & 1.82$_{-0.06}^{+0.06}$ & 1.05(14) & 1.86$_{-0.19}^{+0.20}$ &  1.87$_{- 0.47}^{+ 0.52}$ & 1.12(13)\\
 1334$-$127  & 1.73$_{-0.16}^{+0.15}$ & 1.21(14) & 2.09$_{-0.36}^{+0.18}$ &  6.77$_{- 1.79}^{+ 2.42}$ & 1.05(13)\\
 1351$-$018  & 1.86$_{-0.37}^{+0.30}$ & 1.67(12) & 1.14$_{-0.46}^{+0.51}$ &  0.82$_{- 0.82}^{+ 1.54}$ & 1.20(11)\\
 1510$-$089  & 1.92$_{-0.15}^{+0.15}$ & 0.99(29) & 1.77$_{-0.29}^{+0.29}$ &  6.80$_{- 0.70}^{+ 0.70}$ & 0.97(28)\\
 1519$-$273  & 2.03$_{-0.43}^{+0.38}$ & 1.16(9)  & 2.37$_{-0.93}^{+2.63}$ & 11.84$_{- 6.09}^{+37.32}$ & 1.22(8) \\
 1725$+$044  & 2.01$_{-0.25}^{+0.23}$ & 0.79(11) & 2.63$_{-0.90}^{+1.82}$ & 12.83$_{- 6.60}^{+14.01}$ & 0.72(10)\\
 2126$-$158  & 1.06$_{-0.17}^{+0.15}$ & 1.04(22) & 1.52$_{-0.35}^{+0.55}$ &  8.81$_{- 2.79}^{+ 8.69}$ & 0.77(21)\\
 2128$-$123  & 2.16$_{-0.42}^{+0.42}$ & 0.52(8)  & 2.36$_{-1.16}^{+1.16}$ &  5.60$_{- 4.90}^{+ 4.45}$ & 0.57(7) \\
 2131$-$021  & 2.05$_{-0.47}^{+0.41}$ & 0.41(4)  & 1.86$_{-1.18}^{+1.27}$ &  3.51$_{- 3.24}^{+ 5.12}$ & 0.50(3) \\
 2134$+$004  & 1.44$_{-0.19}^{+0.18}$ & 1.08(14) & 1.72$_{-0.39}^{+0.39}$ &  6.06$_{- 1.89}^{+ 2.75}$ & 1.03(13)\\
 2145$+$067  & 1.47$_{-0.31}^{+0.26}$ & 0.48(11) & 1.69$_{-0.54}^{+0.59}$ &  6.28$_{- 2.71}^{+ 5.52}$ & 0.46(10)\\
 2203$-$188  & 2.28$_{-0.41}^{+0.41}$ & 0.70(12) & 3.00$_{-1.25}^{+1.25}$ &  5.30$_{- 5.30}^{+ 5.70}$ & 0.69(11)\\
 2223$-$052  & 1.73$_{-0.11}^{+0.10}$ & 1.29(39) & 2.03$_{-0.22}^{+0.24}$ &  6.73$_{- 1.10}^{+ 1.28}$ & 1.15(38)\\
 2227$-$399  & 2.15$_{-0.05}^{+0.05}$ & 0.94(51) & 2.10$_{-0.16}^{+0.17}$ &  1.08$_{- 0.33}^{+ 0.38}$ & 0.95(50)\\
 2240$-$260  & 2.15$_{-0.29}^{+0.33}$ & 0.84(5)  & 1.79$_{-0.69}^{+0.42}$ &  0.78$_{- 0.78}^{+ 2.22}$ & 0.87(4) \\
 2344$+$092  & 2.31$_{-0.14}^{+0.12}$ & 0.76(29) & 2.00$_{-0.28}^{+0.29}$ &  3.99$_{- 0.99}^{+ 1.07}$ & 0.66(28)\\
 2351$-$154  & 1.32$_{-0.15}^{+0.14}$ & 1.65(13) & 1.89$_{-0.42}^{+0.43}$ &  4.45$_{- 1.61}^{+ 1.95}$ & 1.31(12)\\
\noalign{\smallskip}\hline                                                                          
\end{tabular}                                                                                       
\end{table}

\section{Results}

\subsection{X-ray detection rates}

163 sources ($\sim$51 per cent) of the complete sample are detected in soft X-rays
at the 3$\sigma$ confidence level. For the remaining 160 objects the 
2$\sigma$ upper limits to the X-ray count rates were determined. If only the
{\em ROSAT} All--Sky Survey data are considered the number of detections is reduced to
151 ($\sim$47 per cent). 

In this section we will discuss how the probability of a source to be 
detected in
X-rays depends on various source parameters. Only the results from the RASS 
are used in this analysis in order to avoid any bias by the usually much 
longer observation times of the pointed {\em ROSAT} observations. 

\begin{figure}
\hspace*{-0.8cm}
\psfig{figure=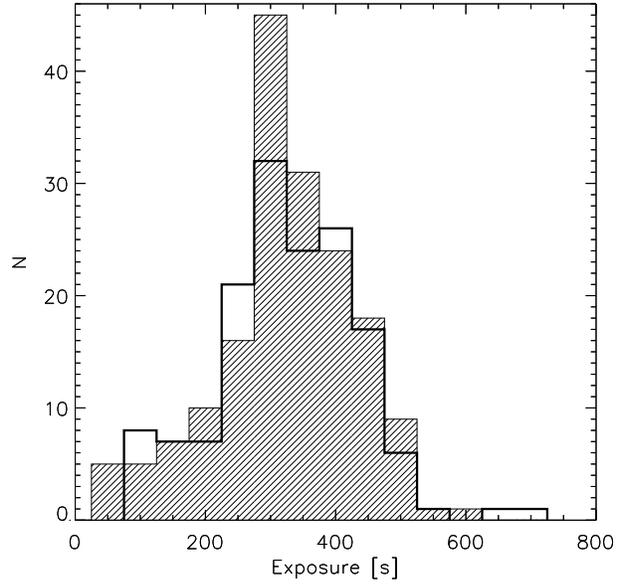,width=9cm}
\caption[]{\label{exposure} Distribution of the effective RASS exposures 
for the sources detected (solid line) and the sources with upper limits 
(hatched). The two distributions are indistinguishable.} 
\end{figure}

To begin with, we note that the distributions of detections and upper limits 
are not significantly different in terms of the RASS exposure. This is 
illustrated in Fig. \ref{exposure}. The RASS exposures range from 65 sec. to 
about 730 sec with a median of 348 sec. A Kolmogorov--Smirnov (KS) test gives 
a probability of $P = 0.996$, i.e. the two distributions are practically 
identical.

\begin{figure}
\hspace*{-0.8cm}
\psfig{figure=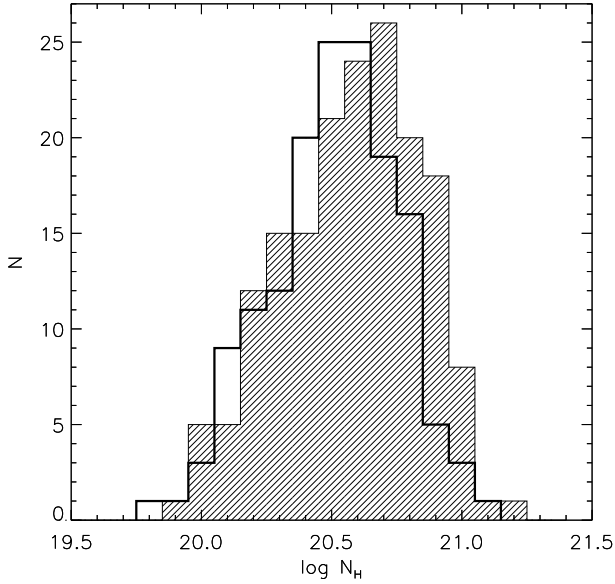,width=9cm}
\caption[]{\label{nh} Histogram of the Galactic $N_{\rm H}$ values for the 
sources
detected in the RASS (solid line) and the sources with upper limits (hatched).
Note the shift towards higher $N_{\rm H}$ values for upper-limit sources.} 
\end{figure}

However, as shown in Fig. \ref{nh}, the distribution of upper limits (hatched) 
is systematically shifted towards higher Galactic $N_{\rm H}$ values. 
A KS-test gives $P = 0.048$, i.e. the hypothesis that the two distributions
are the same can be rejected at a significance level of 4.8 per cent.
A high value for the Galactic absorption certainly reduces the probability 
for a source to be detected in soft X-rays, but the influence is difficult 
to quantify. In any case, this effect is hardly the main reason for a source 
not to be detected in the RASS.

The detection rates for the various optical classifications are $\sim$47.0 per cent 
(110 out of 234) for quasars, 62.9 per cent (22 out of 35) for galaxies and 15.8 per cent
(6 out of 38) for ''faint'' sources.
13 out of the 14 BL Lac objects in the sample are detected in the {\em ROSAT} 
All--Sky Survey. The only exception is PKS 2131-021, a {\it bona fide} BL Lac 
from the 1-Jy sample. It is only detected in a $\sim$8 ksec pointed 
observation. Both currently unidentified objects are not seen in the RASS.

\begin{figure}
\psfig{figure=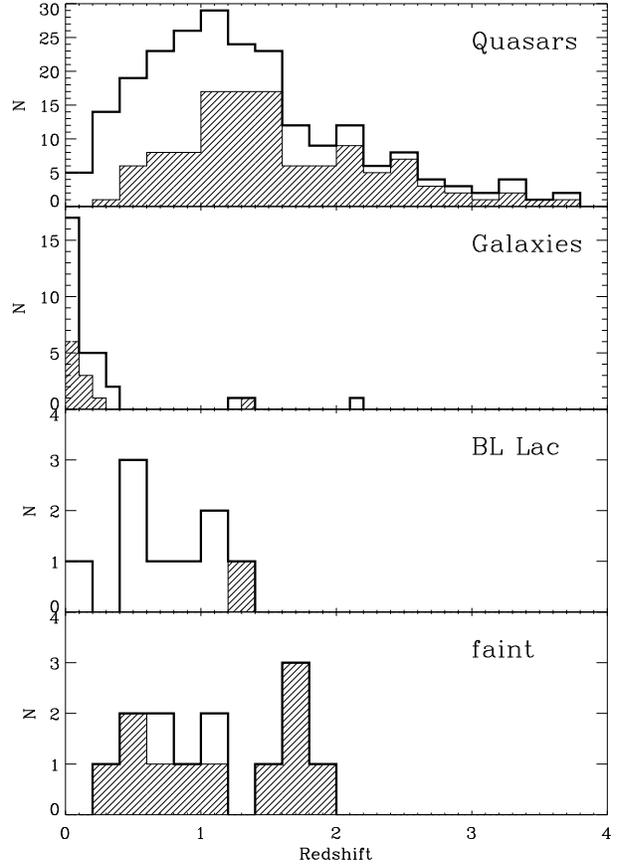,width=9cm}
\caption[]{\label{red} Redshift distributions for the main classes of 
objects in the sample.
The distribution for the {\it total} number of objects in each class (i.e. 
detections plus upper-limits) is drawn with as solid line, whereas the 
upper-limit sources are represented by the hatched areas.} 
\end{figure}

Fig. \ref{red} shows the redshift distributions of the various object classes. 
In each
panel the distribution of the total number of objects, i.e. detections plus 
upper limits, in the respective class is plotted as a solid line. The 
upper limits alone are represented by the hatched area. 

226 quasars have known redshifts and the average redshift is $z = 1.32$. 
Obviously, the fraction of upper limits increases with redshift. The
average redshift of the detected quasars is $z = 1.08$ compared to $z = 1.55$
for the non-detections. A KS-test confirms that the two distributions are
different ($P\approx 1.2\times 10^{\rm -6}$).
Not surprisingly, this suggests that the 
distance of the source is the main parameter influencing the X-ray 
detection probability. A similar trend holds for BL Lac objects, although 
the involved number is small. PKS 2131-021, the only non-detected 
BL Lac object happens to be at the highest redshift ($z = 1.285$). Note that 
the redshift reported by Drinkwater et al. \shortcite{drinkwater} is 
different from the 
(unconfirmed) value published previously ($z = 0.557$; Wills \& Lynds 1978;
Stickel et al. 1991). No trend with redshift is visible for the galaxies and 
the optically 'faint' objects. Comparing the redshift distributions of the 
detected and the undetected objects, the KS-test gives $P = 0.89$
and $P = 0.88$ for galaxies and 'faint' objects, respectively. This indicates 
that properties intrinsic to the sources also influence the detection 
probability, like for example the optical or the radio luminosity. The 
statistical properties of the redshift distributions of all objects classes 
are summarized in Table \ref{zstat}.

The 44 sources without known redshift do not affect our conclusion that 
distance is the main factor which determines the probability for X-ray 
detection. Only 8 out 234 quasars ($\sim$3 per cent) have no redshift. Although 
25 of the 38 'faint' sources have no redshift, the comparison of the 
detection rates for sources with and without redshift shows that these
are identical.  

\begin{table}
\caption[]{\label{zstat}Average and median of the redshift distributions 
of the various
object classes.}
\begin{tabular}{@{}lrccl}
\noalign{\smallskip} \hline \noalign{\smallskip}
\multicolumn{1}{c}{Class} & \multicolumn{1}{c}{N} & Total & Detections & 
\multicolumn{1}{c}{Upper limits} \\
\multicolumn{1}{c}{(1)} & \multicolumn{1}{c}{(2)} & (3) & (4) & 
\multicolumn{1}{c}{(5)}\\ 
\noalign{\smallskip} \hline \noalign{\smallskip}
Quasars & 226 & 1.32 (1.18) & 1.06 (0.91) & 1.56 (1.42)\\
Galaxies & 32 & 0.25 (0.10) & 0.27 (0.10) & 0.21 (0.09)\\
BL Lacs & 9 & 0.74 (0.77) & 0.67 (0.77) & 1.29 (...)\\
'faint' & 13 & 1.11 ( 1.02) & 0.98 (1.18) & 1.13 (1.02)\\
\noalign{\smallskip} \hline \noalign{\smallskip}
\end{tabular}  

\medskip

Column (1) Object class. Column (2) Number of objects with known redshift. 
Column (3) Average redshift of the total number of objects in the respective 
class. The value in brackets gives the median. Columns (4),(5) The same for 
detections and upper limits.
\end{table}

\subsection{X-ray positions}

We compared the X-ray positions derived from the {\em ROSAT} All--Sky Survey 
data with the radio positions of the sample sources. Angular distances up to
$\sim$75 arcsec are found, although the majority of the X-ray sources
lies within 50 arcsec of the radio position, which corresponds to about
2.5$\sigma$ of the average positional accuracy in the RASS \cite{voges}.
The positional differences are mainly due to the X-ray observation, because
the optical/radio positions from the flat-spectrum Parkes sample are known
with arcsec accuracy \cite{drinkwater}. We are therefore confident that the 
X-ray sources detected are truly associated with the radio sources. There is 
one X-ray source at a distance of 164 arcsec from the radio position of 
1509+022. This X-ray source is also included in the {\em ROSAT} Bright 
Source catalog \cite{vogesa}, which is based on the latest reprocessing of 
the Survey data (RASS II). The position derived from these data corresponds 
to within 8 arcsec to the radio source position. We therefore keep 1509+022 
as an X-ray detection, because the large positional difference is obviously 
due to an inaccurate attitude solution of the RASS I data.

\subsection{Variability}

\begin{figure}
\hspace*{-0.8cm}
\psfig{figure=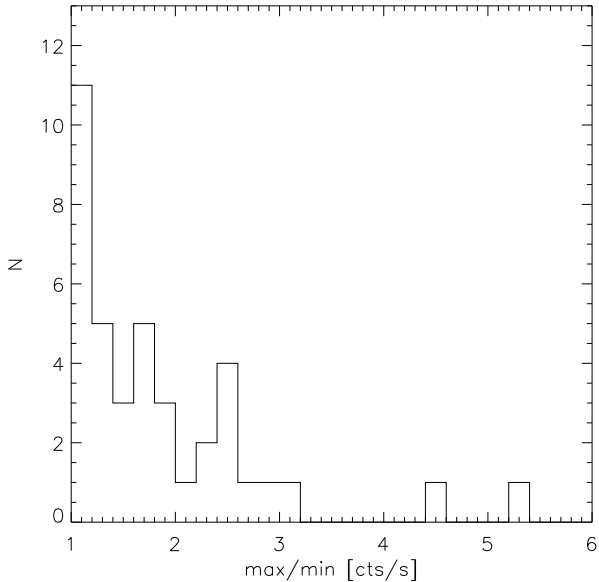,width=9cm}
\caption[]{\label{var}The ratio of the maximum and the minimum measured 
count rate for
objects detected in both the RASS and pointed PSPC observations. For clarity,
the most variable object (PKS 1514-241) is not shown in the diagram. It varied
by more than a factor of 17 within the two observations (see text).} 
\end{figure}

37 sources were detected in the RASS and in one or more {\em ROSAT} PSPC pointings.
We are therefore able to compare the X-ray count rates 
from these observations. Variability is parametrized by the ratio
of the maximum and the minimum observed count rate. The distribution of this 
quantity is shown in Fig. \ref{var}. For about 40 per cent of the sources the observed
variations are below 40 per cent and can thus not be distinguished from statistical
fluctuations. Only 4 sources ($\sim$11 per cent) have varied by more than a factor 
of three
(the quasars 3C 279 and PKS 1510-089 as well as the BL Lac objects 
PKS 1514-241 and PKS 1519-273). As already noted by Siebert et al. 
\shortcite{siebert}, the most extreme object is the BL Lac PKS 1514-241 
(Ap Lib), which 
is clearly detected in the RASS, but not in a 2.8 ksec pointed PSPC 
observation. The upper limit from the latter observation implies that Ap Lib 
varied by more than a factor of 17 within 3 years. 

The distribution in Fig. \ref{var} looks very similar to the corresponding 
distribution
for the bulk of radio-loud quasars as for example described in Brinkmann et al.
(1997a, see their fig. 2). Therefore, within the limits of small number 
statistics, the flat-spectrum Parkes radio sources do not seem to exhibit 
larger X-ray variability than 'ordinary' radio-loud quasars.

\subsection{X-ray spectra}

As described in Section 3.3, we determined the spectral parameters for
an absorbed power-law model for more than one hundred sources of the sample 
either by explicitly fitting the model to the data or by applying the 
hardness ratio technique. In particular the latter method usually results
in rather large uncertainties on the individual parameters and we therefore 
take a statistical approach to investigate the spectral properties of 
flat-spectrum radio sources. 

\begin{table*}
\caption[]{\label{mlgam}Best-fitting parameters for the photon index 
distributions 
from a maximum-likelihood analysis.}
\begin{tabular}{@{}lccc|ccc}
\hline 
      & \multicolumn{3}{c}{Galactic $N_{\rm H}$} 
      & \multicolumn{3}{c}{free $N_{\rm H}$}\\
Class & N & $\Gamma$ & $\sigma$ & N & $\Gamma$ & $\sigma$ \\
\multicolumn{1}{c}{(1)} & (2) & (3) & (4) & (5) & (6) & (7)\\ 
\hline 
quasars & 83 & 1.95$^{\rm +0.13}_{\rm -0.12}$ & 0.37$^{\rm +0.11}_{\rm -0.08}$
   & 69 & 1.95$^{\rm +0.13}_{\rm -0.08}$ & 0.10$^{\rm +0.10}_{\rm -0.10}$\\
BL Lacs & 10 & 2.40$^{\rm +0.12}_{\rm -0.31}$ & 0.23$^{\rm +0.32}_{\rm -0.22}$
   & 10 & 2.55$^{\rm +0.20}_{\rm -0.29}$ & 0.05$^{\rm +0.43}_{\rm -0.05}$\\ 
galaxies & 17 & 1.70$^{\rm +0.23}_{\rm -0.24}$ & 0.34$^{\rm +0.26}_{\rm -0.14}$
   & 15 & 1.95$^{\rm +0.26}_{\rm -0.23}$ & 0.25$^{\rm +0.27}_{\rm -0.13}$\\
\hline 
\end{tabular}  
\vskip0.2cm
\begin{minipage}{14cm}
\small 
Column (1) Object class. Column (2) Number of objects used in the analysis 
with fixed Galactic $N_{\rm H}$. Columns (3),(4) Best-fitting mean $\Gamma$ 
and width $\sigma$ for the underlying parent population. Errors are 90 per cent
confidence. Columns (5)--(7) The same as (1)--(3), but for the distributions 
resulting from the analysis with $N_{\rm H}$ as a free parameter.
\end{minipage}
\end{table*}

\begin{figure}
\psfig{figure=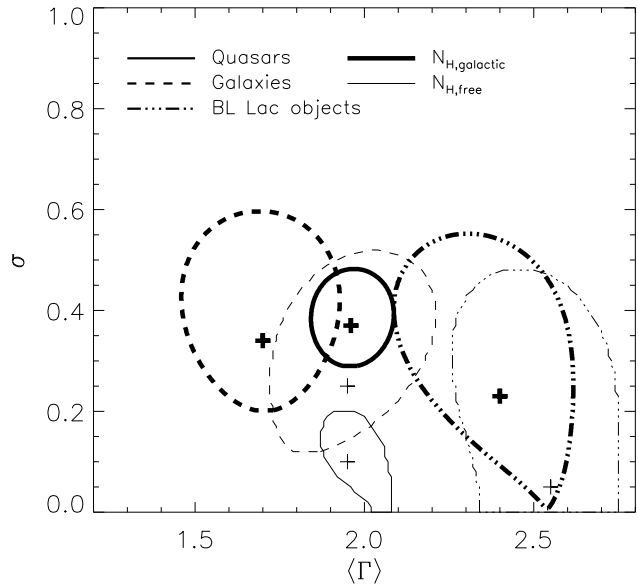,width=9cm}
\caption[]{\label{ml}Result of the maximum-likelihood analysis of the 
distribution of X-ray spectral indices for quasars, galaxies and BL Lac 
objects (mean $\langle\Gamma\rangle$ and intrinsic dispersion $\sigma$) for 
both fixed Galactic $N_{\rm H}$ (thick lines) and free $N_{\rm H}$ (thin
lines). The crosses correspond to the best-fitting parameters, whereas 
the contours represent the 90 per cent confidence ranges.}
\end{figure}

We investigated the resulting photon index distributions for quasars, BL Lac
objects and galaxies using a maximum-likelihood procedure \cite{maccacaro}, 
which gives an estimate of the mean and the 
{\it intrinsic} dispersion
of a supposed parent distribution and which takes into account the errors
on the individual data points. The results of this analysis are presented in
Fig. \ref{ml} and in Table \ref{mlgam}.
 
The average photon index for the flat-spectrum quasars is 
$\langle\Gamma\rangle\approx 1.95^{+0.13}_{-0.12}$ for fixed Galactic 
$N_{\rm H}$ and it remains the same, when absorption is
considered as a free parameter. The soft X-ray spectra of flat-spectrum
radio-loud quasars are thus flatter ($\Delta\Gamma\sim 0.25$) than 
that of 'ordinary' radio-loud quasars ($\langle\Gamma\rangle\sim 2.2$; 
cf. Brinkmann et al. 1995; Brunner et al. 1992) with a similar redshift
range. The latter point is important because of the well-known redshift
dependence of the soft X-ray photon index (Schartel et al. 1996; Brinkmann
et al. 1997a), which will be discussed below.
 
If we only consider sources with $z < 1$ in the maximum-likelihood analysis,
the average spectral index for fixed $N_{\rm H}$ steepens to 
$\langle\Gamma\rangle\sim 2.1\pm 0.1$. This result is completely consistent 
with the results of Brinkmann et al. \shortcite{bys} for an inhomogeneous 
sample of flat-spectrum quasars with 
$z < 1$ from the V\'eron--V\'eron catalog. We thus confirm that the X-ray 
spectrum of flat-spectrum quasars is on average flatter than that of 
steep-spectrum quasars by $\Delta\Gamma\sim 0.1-0.2$. 

\begin{figure}
\psfig{figure=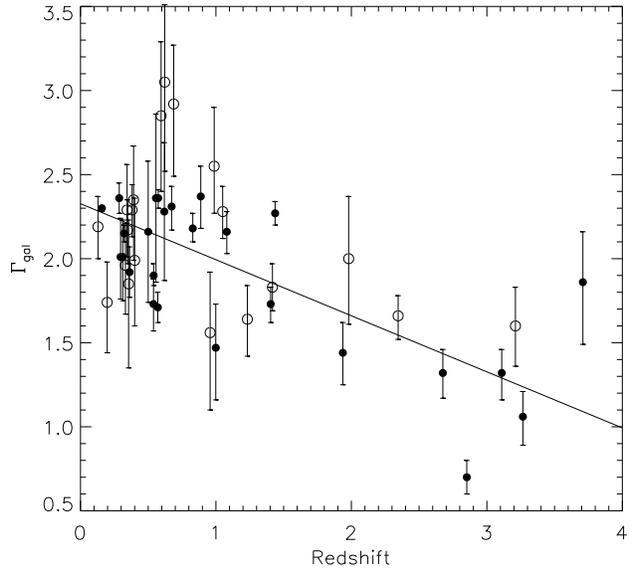,width=9cm}
\caption[]{\label{gz}The photon index $\Gamma$ (for fixed Galactic 
$N_{\rm H}$) as a function of redshift for sources with errors 
$\Delta\Gamma < 0.5$. The straight line indicates the best-fitting regression 
result. Open circles represent sources with the photon index derived from 
hardness ratios, whereas filled circles denote objects with fits to data
from pointed observations.}
\end{figure}

\begin{figure}
\psfig{figure=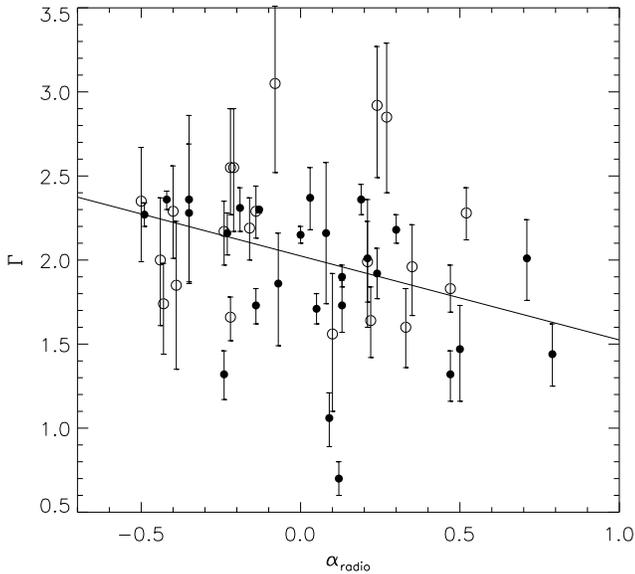,width=9cm}
\caption[]{\label{galp}The photon index $\Gamma$ at fixed $N_{\rm H}$ as a 
function of the radio spectral index $\alpha$ between 2.7\,GHz and 5\,GHz for 
sources with well-determined photon indices ($\Delta\Gamma < 0.5$). The 
straight line indicates the best-fitting regression result. Open circles represent
sources with the photon index derived from the hardness ratios, whereas 
filled circles denote objects with fits to the data.}
\end{figure}

The non-zero dispersion $\sigma$ of the photon index distribution for 
quasars indicates 
that the width of the distribution is not only due to statistical fluctuations,
but that the distribution is intrinsically broadened. This is not 
surprising, since the measured photon index also depends on several other
source parameters like for example redshift (see Fig. \ref{gz}) and radio 
spectral index (see Fig. \ref{galp}). The intrinsic 
dispersion of the distribution is significantly smaller for the fits with
$N_{\rm H,free}$. This might either indicate that a simple absorbed power-law 
is not a 
valid description of the soft X-ray spectrum for a number of sources or that 
the spectra are altered by additional absorption. The latter idea will be 
discussed in detail in Section 5. Here we only note that the quasar with the
flattest X-ray spectrum for fixed Galactic $N_{\rm H}$ is PKS 0438-436 
at $z = 2.85$ ($\Gamma\approx 0.7\pm 0.1$). The fit, however, is statistically 
not acceptable (cf. Table \ref{fits}). The free absorption result indicates 
additional absorption
in the line-of-sight, which seems to be quite common for high redshift 
radio-loud quasars (Cappi et al. 1997; Elvis et al. 1994).
 
The BL Lac objects contained in the sample have significantly steeper soft
X-ray spectra than the other object classes 
($\langle\Gamma\rangle\approx 2.4$).
The results are consistent for the one and the two parameter fits. 
All BL Lacs
are radio-selected and, moreover, belong to the class of objects with 
low-energy cutoffs of the spectral energy distribution \cite{giommip}. Apart 
from one, all BL Lacs for which spectral information could be
obtained are also included in the analysis of the 1-Jy BL Lac sample by Urry
et al. \shortcite{urry96}. Our average photon index is steeper than that of 
the total
1-Jy sample ($\langle\Gamma\rangle\sim 2.2$; Urry et al. 1996), but still
consistent within the errors. 

Radio galaxies exhibit harder X-ray spectra on average compared to quasars
and BL Lac objects. From the 17 galaxies with spectral information
available we find $\langle\Gamma\rangle\approx 1.7$. The result is consistent
with the average spectral indices for much larger samples of radio galaxies  
presented in Brinkmann et al. \shortcite{bsb} and Brinkmann et al. 
\shortcite{bsr}.

The dependence of the quasar soft X-ray photon index on redshift is shown in 
Fig. \ref{gz} for sources with well-determined spectra, i.e. 
$\Delta\Gamma < 0.5$. According to Schartel et al. \shortcite{schartelw}, it 
is due to the fact that the {\em ROSAT} PSPC samples the X-ray spectrum at 
increasingly higher energies in the rest-frame of the source as the redshift 
increases. As a consequence, the steep soft X-ray component is shifted out 
of the PSPC energy range and the generally flatter X-ray component at higher 
energies starts to dominate in the observed 0.1--2.4 keV range. This effect 
is also seen in the present sample of flat-spectrum quasars. A Spearman rank
correlation test gives a probability of $P=0.0035$ that the observed 
correlation occurs by chance. A weighted least squares regression analysis 
gives $\Gamma = (2.33\pm 0.08) - (0.33\pm 0.06)\times z$. The result is
perfectly consistent with the findings of Brinkmann et al. (1997a) for a much 
larger, but inhomogeneous sample of flat-spectrum quasars. Interestingly, the
dependence on redshift is slightly steeper for flat-spectrum quasars than 
for steep-spectrum quasars ($\Gamma = (2.29\pm 0.08) - (0.19\pm 0.11)\times z$.
This may be interpreted in terms of an selection effect. In flux limited 
samples only the most luminous objects are included at any given redshift
(Malmquist-bias). For flat-spectrum radio samples this means that at higher 
redshifts one preferentially selects objects, where the spectral energy 
distribution is dominated by beamed emission from the radio jet. Therefore
it is the combination of the intrinsically higher energies sampled with the
PSPC observations and the increasing contribution of the flat X-ray component 
from the jet that produce a steeper gradient of the photon index with 
redshift compared to steep-spectrum quasars where the latter effect is
less important.

\begin{table*}
\caption[]{\label{corr} Results of the correlation and regression analysis.}
\begin{tabular}{lccccccccccc}
\noalign{\smallskip} \hline \noalign{\smallskip}
  &  & N & $N_{\rm UL}$ & $\tau$ & P$_{\rm 12}$ & a$_0$ & a$_1$ & a$_0^{'}$ & 
    a$_1^{'}$ & $\tau_{\rm xy.z}$ & P$_{\rm 12.3}$\\
\multicolumn{1}{c}{(1)} & (2) & (3) & (4) & (5) & (6) & (7) 
 & (8) & (9) & (10) & (11) & (12)\\ 
\noalign{\smallskip} \hline \noalign{\smallskip}
quasars  & L$_r$-L$_x$ & 226 & 110 & 0.22 & $<$10$^{\rm -7}$ & 22.7$\pm$2.4 & 0.66$\pm$0.07 & 18.1$\pm$2.1 & 0.80$\pm$0.06 & 0.16 & 9.6$\cdot$10$^{\rm -7}$\\
         & L$_o$-L$_x$ & 225 & 109 & 0.22 & $<$10$^{\rm -7}$ & 29.7$\pm$1.9 & 0.50$\pm$0.06 & 26.1$\pm$1.8 & 0.63$\pm$0.06 & 0.18 & 1.2$\cdot$10$^{\rm -7}$ \\
galaxies & L$_r$-L$_x$ & 33 & 9 & 0.45 & 0.8$\cdot$10$^{\rm -4}$ & 14.4$\pm$4.5 & 0.90$\pm$0.14 & 11.9$\pm$2.9 & 0.99$\pm$0.09 & 0.22 & 0.048\\
     & L$_o$-L$_x$ & 33 & 9 & 0.16 & 0.16 & ... & ... & ... & ... & ... & ...\\
BL Lacs & L$_r$-L$_x$ & 10 & 0 & 0.67 & 0.012 & 13.7$\pm$5.4 & 0.93$\pm$0.16 & ... & ... & 0.41 & 0.276\\
        & L$_o$-L$_x$ & 10 & 0 & 0.78 & 0.4$\cdot$10$^{\rm -2}$ & 14.8$\pm$4.3 & 0.99$\pm$0.14 & ... & ... & 0.61 & 0.201\\
\noalign{\smallskip} \hline \noalign{\smallskip}
\end{tabular}  
\vskip0.2cm
\begin{minipage}{17cm}
\sloppy
Column (1) Object class. Column (2) Type of correlation (independent -- 
dependent variable). Column (3) Number of sources used. Column (4) Number of 
upper limits. Column (5) Kendall's $\tau$ correlation coefficient. Column (6) 
Probability that the observed correlation occurs by chance from intrinsically 
uncorrelated data according to a Kendall's $\tau$ test. Columns (7),(8) 
Parameters of the regression line $y = a_{\rm 0} + a_{\rm 1} \cdot x$. 
Columns (9),(10) Same as (7) and (8), but only detected sources are taken 
into account. Column (11) Partial correlation coefficient, with redshift 
effects eliminated. Column (12) Significance of the partial correlation coefficient.
\end{minipage}   
\end{table*}

As already mentioned, the photon indices of flat-spectrum quasars are also 
correlated with the radio spectral index between 2.7\,GHz and 5\,GHz. This 
correlation is illustrated by Fig. \ref{galp} for quasars with well-determined 
photon indices ($\Delta\Gamma < 0.5$). The flatter or the more inverted the 
radio continuum is, the harder is the soft X-ray spectrum. The correlation 
holds for photon indices determined at fixed Galactic $N_{\rm H}$ as well 
as for those determined with free absorption. A Spearman rank correlation
analysis gives a probability of $P=0.040$ that the observed correlation
occurs by chance. A weighted least squares regression analysis gives 
$\Gamma = (2.02\pm 0.05) - (0.50\pm 0.19)\times\alpha$. We note that
Brinkmann et al. (1997a) find no significant correlation of $\Gamma$ with
$\alpha$ for flat-spectrum quasars, which may be due to the fact that we
constrain our analysis to objects with well-determined photon indices
(almost half of them from fits to data obtained in pointed PSPC observations). 
 
Within the framework of orientation dependent unification scenarios for 
radio-loud AGN (Urry \& Padovani 1995 and references therein), steep- and 
flat-spectrum radio sources are the increasingly aligned counterparts of 
Fanaroff-Riley type II radio galaxies \cite{fanaroff}. Therefore the radio 
spectral index roughly indicates the orientation of the source with respect 
to the line of sight. In this model the observed correlation between X-ray 
and radio spectral index can be interpreted as being due to a beamed X-ray 
component with a harder spectrum, which is directly related to the radio 
emission from the radio core and which dominates the total X-ray 
emission at small angles to the line-of-sight.     
    
\subsection{Luminosity correlations}

\begin{figure}
\hspace*{0.cm}
\psfig{figure=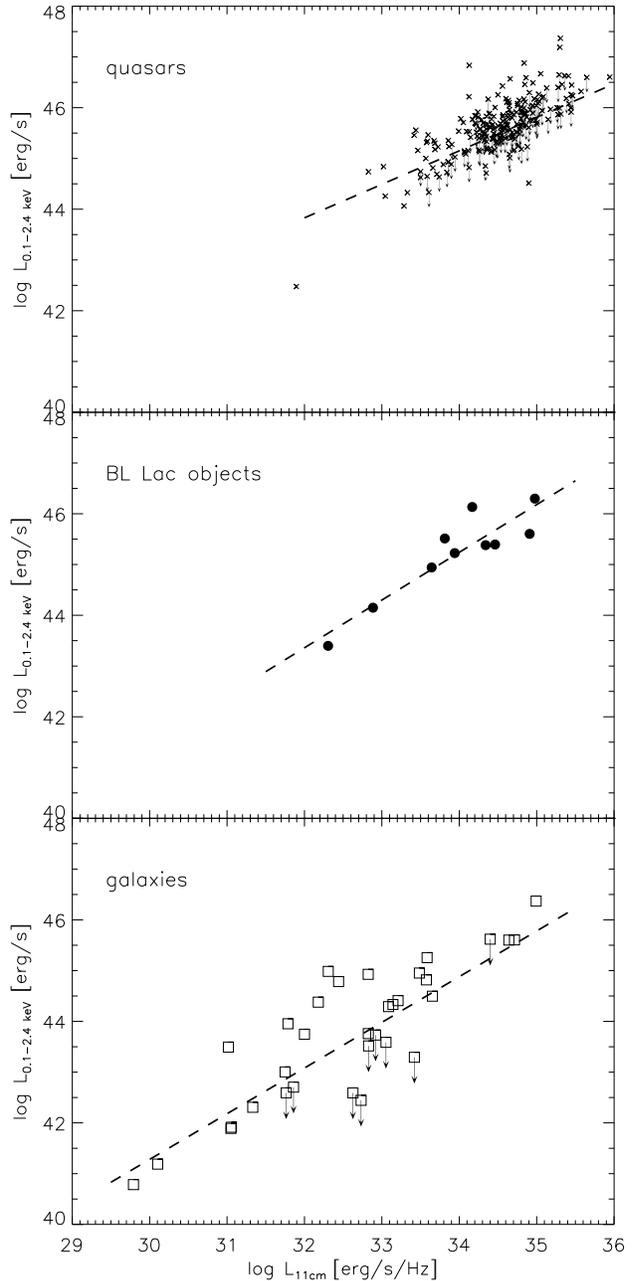,width=9cm}
\caption[]{\label{lxlr}Soft X-ray luminosity of quasars, BL Lac objects 
and galaxies as a function of the total 2.7\,GHz (11\,cm) radio luminosity. 
The results of the regression analysis including the upper limits are 
indicated by the dashed lines (cf. Table \ref{corr}).}
\end{figure}

To investigate whether the emission at two different frequencies is {\em 
intrinsically} related, we seek for correlations in luminosity-luminosity
diagrams. It is often stated that observed luminosity correlations are 
significantly affected by redshift effects, i.e. the correlation appears 
in absolute luminosities even for uncorrelated fluxes, because the (narrow) 
flux range is stretched out by a large distance factor. Feigelson \& Berg 
\shortcite{feigelson} invert this argument by pointing out that objects will 
not be detected 
in small flux ranges at two frequencies {\em unless} the luminosities are 
intrinsically related. Generally, the more distant objects will exhibit flux 
upper limits in one band if no relationship exists. If these upper limits are 
properly taken into account, intrinsic relationships can be recovered from 
luminosity-luminosity diagrams \cite{feigelson}. We further note 
that intrinsic relationships cannot be determined from flux-flux diagrams 
anyway, unless the underlying relationship is linear (e.g. Kembhavi, 
Feigelson \& Singh 1986). 
   
The radio, optical and X-ray luminosities of our sample were calculated 
assuming $H_{\rm 0} = 50$ km s$^{\rm -1}$ Mpc$^{\rm -1}$ and $q_{\rm 0} = 0.5$.
For the K-correction we used in the radio band the radio spectral indices 
between 2.7\,GHz and 5\,GHz as given in Drinkwater et al. \shortcite{drinkwater}.
For the
X-ray luminosity the photon indices for Galactic absorption as derived in 
Section 3.3 were used if the error is smaller than 0.5. Otherwise the average 
values for quasars, galaxies and BL Lac objects were applied. The optical
luminosity is corrected for Galactic reddening and K-corrected to 
4400 \AA\ with a spectral index of $\alpha = -0.5$.

Correlation and regression analyses were performed with {\sevensize ASURV}
(Rev.1.3; La~Valley, Isobe \& Feigelson 1992), which is particularly designed 
for censored data.
For the correlation analysis we applied the modified Kendall's $\tau$ and for
the regression analysis we used the parametric EM algorithm from this software 
package. To investigate the influence of redshift on the correlations we 
performed a partial correlation analysis using the procedure described in 
Akritas \& Siebert \shortcite{akritas}, which allows us to determine the 
partial 
correlation coefficient in the presence of upper limits and to calculate 
the significance of this corrected correlation coefficient. The results from 
this analysis are summarized in Table \ref{corr}. For completeness we give 
the parameters for both the $L_{\rm o}-L_{\rm x}$ and the 
$L_{\rm r}-L_{\rm x}$ correlation although we will discuss only the 
latter in detail. 
   
In Fig. \ref{lxlr} we show the relation between radio and X-ray luminosity
for quasars, galaxies and BL Lac objects. For all three object classes the
X-ray and the radio luminosities are obviously correlated with each other.
The slope of the regression line for quasars is $a_{\rm 1} = 0.61\pm 0.08$, 
which is consistent with the findings of Brinkmann et al. \shortcite{bys} for 
the 
inhomogeneous sample of X-ray detected flat-spectrum quasars from the 
V\'eron-V\'eron catalog. They find $a_{\rm 1} = 0.72\pm 0.11$ using orthogonal 
distance regression, which is superior to simple least squares methods
for theoretical reasons, but which is unfortunately not available for 
censored data. We note that neglecting the upper limits only slightly affects 
the slope of the correlation.
The partial correlation analysis confirms that the observed relation is not 
due to a common dependence of both luminosities on redshift. Although the 
probability for the observed correlation to occur by chance increases 
if redshift effects are taken into account, the correlation is still 
highly significant ($P = 9.6\times 10^{\rm -7}$).   

Compared to quasars, the correlation between radio and X-ray luminosity
seems to be steeper for galaxies and BL Lac objects. In the case of galaxies,
the slope of the correlation is consistent with previous findings within
the mutual uncertainties (Fabbiano et al. 1984; Brinkmann et al. 1995).
However, an interpretation of the relation between total radio and X-ray 
luminosity 
in terms of nuclear activity is hampered by contaminating X-ray emission 
from various galactic processes or surrounding clusters of galaxies in many
of the radio galaxies. The spatial resolution in the {\em ROSAT}
All--Sky Survey generally does not allow to isolate the AGN contribution to 
the total X-ray emission.

The slope of the regression line for the BL Lac objects is consistent
with unity, which might argue for a direct physical relation between the radio
and the X-ray emission. Since the radiation in both wavebands is thought to be
dominated by the relativistic jet (e.g. Kollgaard 1994), this would not be 
surprising. However, the partial correlation analysis indicates that the 
correlation for BL Lac objects is strongly influenced by selection effects 
and is no longer significant, if redshift effects are taken into account.
On the other hand we note that the number of objects is rather low. Therefore 
the results of statistical analyses have to be interpreted with caution.

\section{Is there evidence for dust reddened quasars?}

Based on the large scatter in the $B_{\rm J}-K$ colors of the quasars
in the present sample, Webster et al. \shortcite{webster} claimed that
there might be a large population of previously undetected, red
quasars. They suggested that this reddening is due to dust intrinsic to 
the quasars. In this section we will investigate if there are any arguments
from the X-ray data in support of this hypothesis. A discussion of the results
will be presented in Section 6.

\subsection{Dust extinction and expected X-ray absorption}

If the optical reddening is due to dust, it is expected that the reddest 
quasars are also intrinsically absorbed in soft X-rays. Using Galactic 
gas-to-dust ratios (i.e. 
$N_{\rm HI+H_2} = 5.8 \times 10^{\rm 21} \cdot E_{\rm B-V}$ cm$^{\rm -2}$; 
Bohlin, Savage \& Drake 
1978), the extinction coefficients $R_{\lambda} = A_{\lambda}/E_{\rm B-V}$ 
from Seaton \shortcite{seaton} and the energy dependent X-ray absorption 
coefficients given by Morrison \& McCammon \shortcite{morrison}, we can 
estimate the influence of the absorption by cold gas associated with the dust
on the soft X-ray flux for a given {\em observed} extinction $A_{\rm B,obs}$. 
For example, $A_{\rm B,obs} = 5$ corresponds to an intrinsic $N_{\rm H}$ of
$\sim 7\times 10^{21}$ cm$^{\rm -2}$ for local objects, which reduces the 
{\em observed} X-ray flux in the total 0.1--2.4 keV energy band by more than a 
factor of three. For redshifts of $z \approx 1.3$, which is the average 
redshift of the present quasar sample, an observed optical extinction of
$A_{\rm B,obs} = 5$ corresponds to a rest-frame $N_{\rm H}$ of 
$\sim 3\times 10^{21}$ cm$^{\rm -2}$, which reduces the observed 0.1--2.4 
keV flux still by a factor of $\sim1.4$.

We further note that the decrease of the observed optical (4400{\AA}) flux
by dust extinction, which is given by $\approx 10^{-0.4\cdot A_{\rm B}}$, 
generally is larger than the corresponding decrease of the soft X-ray flux 
caused by absorption by cold gas associated with the dust, in particular 
for higher redshifts. For example, an {\em intrinsic} extinction of 
$A_{\rm B} = 2$ at a redshift of $z = 1$ reduces the {\em observed} 4400{\AA} 
flux by a factor of 50, whereas the {\em observed} 0.1--2.4 keV flux is only
reduced by a factor of $\approx 1.3$. This point is important in the 
discussion of the $\log (f_{\rm x}/f_{\rm o})$ distribution in Section 5.5.    

\subsection{Detection probability}

\begin{figure}
\psfig{figure=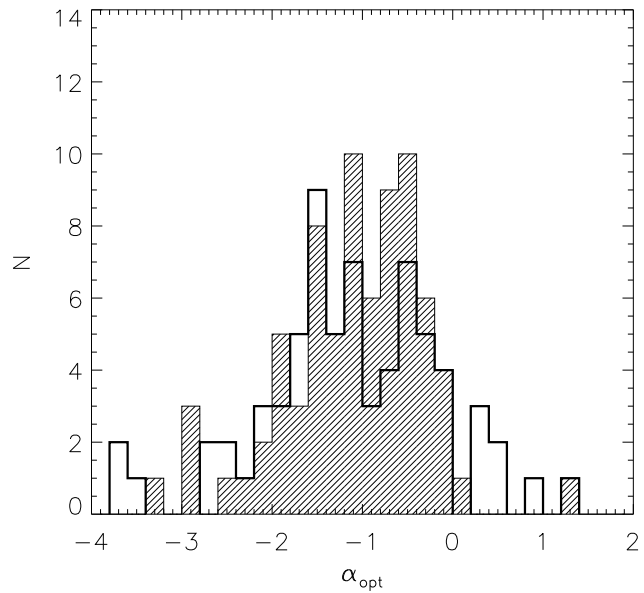,width=9cm}
\caption[]{\label{aopt} Comparison of the power-law index of the optical
continuum for X-ray detections (solid line) and upper limits (hatched).
The two distributions are statistically indistinguishable.}
\end{figure}

In a first step we compared the shape of the optical continuum for 
the detected quasars with those for the upper-limit sources. The result 
is shown in Fig.~\ref{aopt} for the total sample. The optical continuum is 
parametrized by the power-law slope $\alpha_{opt}$ (f$_{\nu}\propto 
\nu^{\alpha}$), which is given in Francis et al. (in prep.) for 181 
sources of the sample. They find a correlation between the optical slopes 
and the $B_{\rm J}-K$ colors. The optical slope can therefore be used to 
parametrize 
the redness of the optical continuum, which in turn is a measure of the
amount of dust present in the source, if one follows the interpretation
of Webster et al. \shortcite{webster}. 

The values of $\alpha_{\rm opt}$ show a large scatter and range from 
$-4$ to $+1.5$, but nevertheless the distributions for detections and 
upper limits look very similar. 
In particular, the upper limits do not cluster towards 
red optical continua. This is confirmed by a Kolmogorov--Smirnov test, 
which gives $P = 0.83$, i.e. if we reject the null-hypothesis that 
the two distributions are drawn from the same parent population, the 
error probability is 83 per cent. Confining the sample to low redshifts ($z < 1.0$),
where the absorption effect on the X-ray flux should be most easily 
detectable, this probability is even higher (97.3 per cent). We therefore conclude 
that intrinsic absorption by gas/dust is not the main reason for a source 
not to be detected in soft X-rays and confirm our conclusion in Section 4.1
that distance is the most important parameter.

\subsection{X-ray luminosities}

\begin{figure}
\psfig{figure=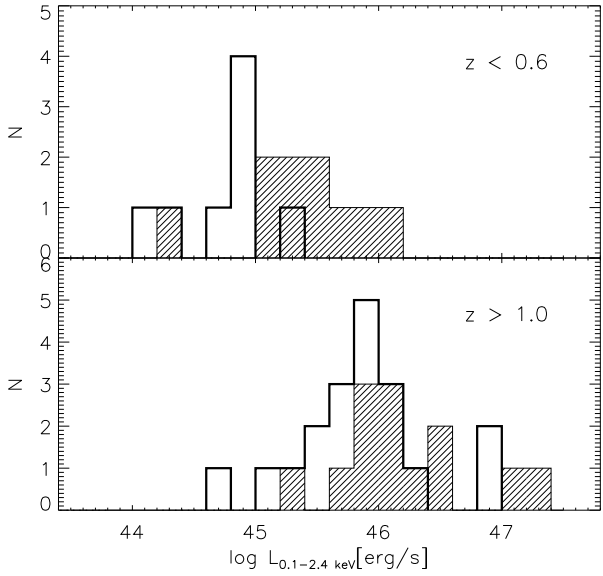,width=9cm}
\caption[]{\label{histo_xlum} The soft X-ray luminosity distributions for 
detected sources with $\alpha_{\rm opt} > -1$ (hatched) and 
$\alpha_{\rm opt} < -1$ (thick line). In the upper panel only low redshift 
'stellar' sources are considered ($z < 0.6$), whereas the high redshift 
sources ($z > 1$) are shown in the lower panel.}
\end{figure}

As shown in Section 5.1, absorption by cold gas associated with the 
inferred amounts of dust can have a significant effect on the soft X-ray
flux and hence the observed X-ray luminosity. In Fig.~\ref{histo_xlum} we 
compare the soft X-ray luminosity distributions
for 'stellar' sources with an optical continuum slope $\alpha_{\rm opt} > -1$ 
(hatched) to those with $\alpha_{\rm opt} < -1$ (thick line). For clarity, only
the detected sources are shown. In the statistical analysis, however, we
included the upper limits. 

For low redshift sources the luminosity distributions are clearly 
different (upper panel). On average, the optically 'red' sources have 
X-ray luminosities which are a factor of 2--3 lower than those of their 'blue' 
counterparts. Using Gehan's generalized Wilcoxon test as implemented in
{\sevensize ASURV}, we get $P = 0.0082$, i.e. the probability for erroneously
rejecting the null hypothesis that the distributions are the same is only 
0.8 per cent. We further note that the redshift distributions of the low redshift 
'red' and 'blue' subsamples are indistinguishable ($P = 0.42$). The effect 
almost completely disappears for sources at higher redshifts ($z > 1$; 
lower panel). The soft X-ray luminosity distributions for the two object 
classes are statistically indistinguishable ($P = 0.69$). 
  
This kind of redshift dependence is the expected signature of intrinsic X-ray 
absorption by cold gas associated with the dust inferred from the observed 
optical reddening. In Section 5.1 we have shown that the effect of the
deduced amount of cold gas on the soft X-ray flux decreases with redshift. 

However, there might be selection effects at work, which
could lead to a similar result. In particular, the low-redshift sources,
although all of them appear stellar on the UKST and POSS I plates, might 
still be contaminated by galaxies or BL Lac objects. Galaxies and BL Lac 
objects are known to have redder optical continua and lower X-ray luminosities
than quasars (Brinkmann et al. 1995). We therefore checked databases,
literature and our own optical data, if any of the low redshift 'red' sources 
actually have to be classified as a galaxy or a BL Lac object. First of 
all, no source shows a strong galaxy continuum in the spectrum as
presented in Francis et al. (in prep.). Only for three sources the 
identification seems questionable according to NED. The sources 1020-103 and 
1034-293 are classified as BL Lacs. However, strong emission lines have been 
reported for these sources (Falomo, Scarpa \& Bersanelli 1994; Falomo 1996). 
These 
sources should therefore be considered as quasars instead of BL Lac objects. 
The third source, 0003-066, is listed as a radio galaxy and BL Lac candidate 
in NED, but Stickel et al. \shortcite{stickel94} classify this source as a 
quasar based on several weak emission lines. 

We conclude that the 
contamination of the low-redshift 'red' sources by low-luminosity galaxies 
and BL Lac objects cannot explain the observed difference in the soft X-ray 
flux distributions.          

\subsection{Soft X-ray spectra}

\begin{figure}
\psfig{figure=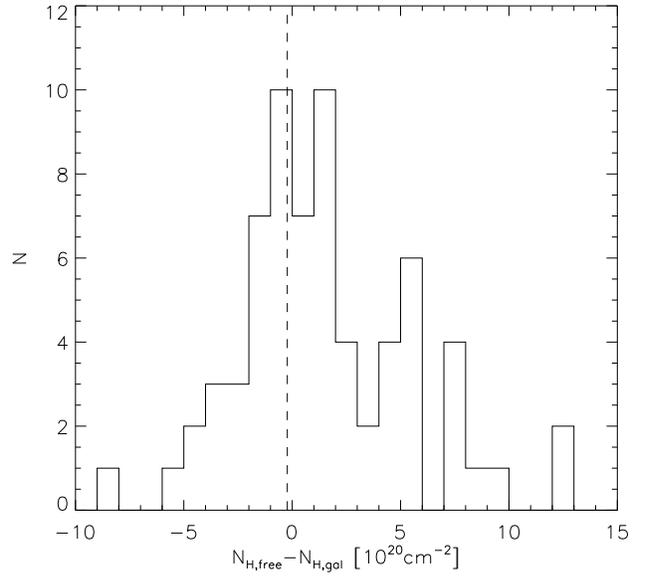,width=9cm}
\caption[]{\label{delnh} Histogram of the difference between the X-ray
derived $N_{\rm H}$ value from the two parameter fits and the Galactic 
$N_{\rm H}$ value. The weighted average (dashed line) is consistent with 
zero excess absorption.}
\end{figure}

Next we investigate if the influence of dust associated absorption can be
seen in the soft X-ray spectra of the sources directly. If excess
absorption exists, it should lead to a substantially higher value of 
$N_{\rm H}$ than the Galactic one in the two parameter fits. Again, the 
effect should be strongest at low redshifts, since the required amounts of 
intrinsic $N_{\rm H}$ to produce an observed extinction of $A_{\rm B}$ may be 
too low to be significantly detected in high redshift sources. In 
Fig.~\ref{delnh} we show the distribution of the difference 
$\Delta N_{\rm H} = N_{\rm H,free} - N_{\rm H,gal}$ between the derived 
$N_{\rm H}$ and the Galactic $N_{\rm H}$ value for the quasars in the sample. 
There seems to be an asymmetry towards positive $\Delta N_{\rm H}$ values. 
However, a more detailed analysis shows that most of the high 
$\Delta N_{\rm H}$ values come from spectral parameters derived from 
hardness-ratios and the error bars are very large. In addition, it is known 
that the hardness ratio technique leads to systematically higher $N_{\rm H}$ 
values for weak sources \cite{yuan}. The weighted average of the 
$\Delta N_{\rm H}$ distribution is 
$\langle \Delta N_{\rm H} \rangle = (-0.22\pm0.68)\times 10^{\rm 20}$ 
cm$^{-2}$, which indicates that excess absorption is not a general feature. 
In particular, the sources with very red optical continua, show no evidence 
for systematically higher absorption in soft X-rays. 

Only six sources show significantly higher observed $N_{\rm H,free}$ compared 
to the Galactic $N_{\rm H}$ value (0438$-$436, 1148$-$001, 1402$+$044, 
2126$-$158, 2223$-$052, 2351$-$154). All of them have high 
redshifts between 1.40 $< z <$ 3.27. A detailed
analysis of the excess X-ray absorption has already been published for
0438$-$436 \cite{cappi}, 2126$-$158 \cite{elvis} and 2351$-$154
\cite{schartelk}. The required amounts of intrinsic $N_{\rm H}$ range 
from $1.7\times 10^{\rm 21}$ to $1.2\times 10^{\rm 22}$ for these sources. 
If we again assume Galactic gas-to-dust ratios, the associated dust extinction
in the (observed) B-band would be much higher than the maximum value 
($\approx 5$ mag) observed by Webster et al. \shortcite{webster}. For
three of these sources (1402+044, 2126-158, 2351-154) optical continuum
slopes are available and they do not indicate a particularly red continuum
($\alpha_{\rm opt} =$ -1.438, -0.588 and -0.297, respectively). Several
explanations for this contradictory result are conceivable and they are
briefly discussed in Section 6.
  
\subsection{$\log(f_{\rm x}/f_{\rm o})$ distribution}  

It has been argued by Webster et al. \shortcite{webster} that the large 
dispersion in the X-ray to optical flux ratios for X-ray selected quasars 
found by Stocke et al. \shortcite{stocke} is indicative of variable dust 
extinction. This has been criticized by Boyle \& di Matteo \shortcite{boyle}, 
who show that the observed scatter of the X-ray to optical flux ratio in their
sample of X-ray selected quasars is much smaller than expected if variable 
extinction is important\footnote{The authors base their analysis on the 
absorption coefficients in the optical and X-ray band, thereby neglecting
the different column densities of absorbing material in the two wavebands,
i.e. dust in the optical and neutral gas in X-rays. The difference is 
important, in particular at low redshifts (cf. Section 5.1)}.  

\begin{figure}
\psfig{figure=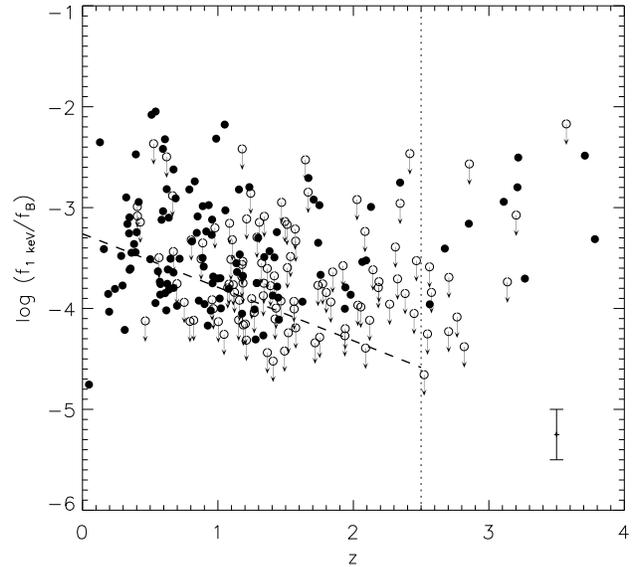,width=9cm}
\caption[]{\label{aox_z}The flux ratio log ($f_{\rm x}/f_{\rm o}$) as a 
function of redshift $z$ for quasars. Open circles plus arrows indicate X-ray 
upper limits. The solid line represents the best regression line for the 
whole sample including the upper limits, whereas the dashed line is for 
$z < 2.5$ objects only. The vertical dotted line is drawn at $z$ = 2.5 for 
illustrative purposes. In the lower right corner we show a conservatively
estimated typical error bar.}
\end{figure}

In Fig. \ref{aox_z} we show the flux ratio log ($f_{\rm x}/f_{\rm o}$) as a 
function of redshift $z$ for the present sample. $f_{\rm x}$ is the 
monochromatic X-ray flux at 1 keV in \ergs Hz$^{-1}$, which is derived from 
the flux 
in the total {\em ROSAT} energy band assuming the average spectral parameters 
derived in Section 4.4. $f_{\rm o}$ is the observed flux at 4400{\AA}, 
calculated from the $B_{\rm J}$ magnitude and corrected for Galactic 
reddening according to the formula (Giommi, Ansari \& Micol 1995)
\[
f_o = 10^{-0.4\times B_{\rm J} + 0.08/\sin(|bII|) - 19.377} \quad ,
\]
where $bII$ is the Galactic latitude of the source. In the lower right 
corner of Fig. \ref{aox_z} we show a conservatively calculated error bar,
assuming a 25 per cent error of the X-ray flux and an uncertainty of the optical
magnitudes of $\Delta m = 0.5$ \cite{drinkwater}. Since the optical and the 
X-ray data were not taken simultaneously, we also included the expected
B-band variability of $\sigma(B) = 0.3$ mag for a typical quasar of our sample
and the rest-frame $\sim 20$ year baseline between the optical and the X-ray 
measurements (Hook et al. 1994).

A decrease of log ($f_{\rm x}/f_{\rm o}$) with redshift up to $z\sim 2.5$
is apparent from Fig. \ref{aox_z} and it turns out to be highly significant 
using Spearman's $\rho$ correlation test as implemented in {\sevensize ASURV} 
($\rho_{\rm sp} = -0.40$ for 204 data points). However, this trend is
mainly caused by the different spectral indices at optical and X-ray energies,
which causes a spurious correlation of the form 
$\log (f_{\rm x}/f_{\rm o}) \propto (1+z)^{-\alpha_{\rm x} + \alpha_{\rm o}}$.
Indeed, using {\sevensize ASURV} to determine the best regression line, we get
$\log (f_{\rm x}/f_{\rm o}) = (-3.26\pm0.12) - (0.53\pm0.10)\times z$, if 
we only consider objects with $z < 2.5$. The slope is roughly 
consistent with the difference in the average spectral indices in the X-ray 
and the optical regime. 
 
Interestingly, the trend with redshift seems to be inverted at redshifts 
greater than $\sim$2.5, which is mainly mimicked by a deficit of high redshift
sources with low $\log (f_{\rm x}/f_{\rm o})$ values\footnote{Note that the 
quasar with the highest upper limit on $\log (f_{\rm x}/f_{\rm o})$ (PKS 
2215$+$02, $z = 3.572$) has been detected in a recent pointed {\em ROSAT} 
HRI observation \cite{siebertb}. The deduced X-ray flux is only slightly 
below the quoted upper limit from the RASS.}. 

Most likely, various effects combine to result in an apparent deficit of
sources with low values of $\log (f_{\rm x}/f_{\rm o})$ at higher redshifts.
Firstly, the X-ray emission could be enhanced, for example by relativistic 
beaming. This hypothesis is supported by the generally very flat or inverted 
radio spectra of these objects and their high core dominance. Basically this 
is a selection effect of the radio sample. At high redshifts only the most 
powerful (and hence probably beamed) objects make it into the flux limited 
sample. Further, the optical flux may be reduced, because the Ly$\alpha$-edge 
shifts into the B-band for $z > 2.5$. Finally we note that a positive 
correlation is also expected in the case of dust extinction, because the
effect of dust on the optical continuum of high redshift sources is much 
stronger than that of the associated neutral gas on the X-ray emission (cf.
Section 5.1). Some of these high redshift objects do show significant intrinsic
absorption in X-rays (see previous section) and they are optically 
dull with $B_{\rm J} > 21$ mag. Therefore it cannot be excluded that their 
high X-ray-to-optical flux ratios are at least in part due to dust extinction. 

Analogous to Boyle \& di Matteo \shortcite{boyle} we investigate the 
dispersion in the distribution of the flux ratio log ($f_{\rm x}/f_{\rm o}$), 
which can be used to at least derive upper limits on the amount of dust 
obscuration. As has been shown in Section 5.1, the effect of intrinsic gas and 
dust on the observed optical flux is much stronger than on the X-ray flux, in 
particular at higher redshifts. To first order, the observed spread in 
log ($f_{\rm x}/f_{\rm o}$) might therefore be almost completely attributed  
to variations in $f_{\rm o}$ due to dust extinction.  

If we apply the Detections-and-Bounds (DB) method of Avni et al. 
\shortcite{avni} in order to properly account for upper limits, we get 
$\langle \log (f_{\rm x}/f_{\rm o}) \rangle = -3.87$ and 
$\sigma[\log (f_{\rm x}/f_{\rm o})] = 0.70$, which is significantly larger 
than the Boyle \& di Matteo value ($\sigma\sim 0.4$). The derived 
$\pm 1\sigma$ width of the $\log (f_{\rm x}/f_{\rm o})$ distribution
corresponds to a variation in $f_{\rm x}/f_{\rm o}$ by a factor of $\sim 25$.
Assuming that all of the scattering is due to dust extinction (and thus 
neglecting other effects like X-ray variability, measurements errors etc.), 
this gives an upper limit to the amount of reddening in the observed B band of
$<$ 3.5 mag. This is slightly below the maximum values claimed by Webster et 
al. \shortcite{webster}, but since the estimate is based on the FWHM of the 
$\log (f_{\rm x}/f_{\rm o})$ distribution, the observed maximum value of 
$A_{\rm B} = 5$ might still be consistent with the observed scatter. 

Clearly, the observed scatter in $\log (f_{\rm x}/f_{\rm o})$ for the
Parkes quasars cannot be regarded as clear evidence for the presence
of dust. We may conclude, however, that it also does not rule out the 
dust extinction hypothesis, as it has been claimed from the corresponding
analysis of X-ray and optically selected samples, which exhibited a much 
smaller scatter in $\log (f_{\rm x}/f_{\rm o})$ (Boyle \& di Matteo 1995).

\section{Discussion}

One of the main aims of this study was to look for any evidence from
the X-ray data in favor of or against the hypothesis of Webster et al. (1995)
that the observed red continua of many of the Parkes flat-spectrum quasars
are due to dust extinction. Our results can be summarized as follows.

First of all, none of the observed X-ray properties is in contradiction
to the dust hypothesis. On the other hand, it is also difficult to prove the
presence of dust from the X-ray data, because the argument is always an 
indirect one. 

The strongest indication for dust comes from the significantly lower
X-ray luminosities of the low-redshift 'stellar' objects with red optical
continua as compared to their blue counterparts. The difference vanishes
for higher redshifts. This is expected, if neutral gas associated
with the dust is responsible for the soft X-ray absorption and the 
corresponding reduction of the X-ray flux, because, for high redshifts, 
the X-ray observation samples increasingly higher energies in the rest-frame 
spectrum of the source and the column density of neutral gas associated 
with the dust inferred from optical reddening is too low to significantly 
affect the X-ray flux at these relatively high rest-frame energies. Therefore 
no difference in the luminosity distributions for red and blue objects is 
measurable any more. For the low redshift sources, the observed difference 
in the average X-ray luminosity is about a factor of two, which is consistent 
with the optical data. We have further shown that the observed luminosity
difference is not due to a contamination of the 'stellar' sources of our sample
with low luminosity 'red' sources like BL Lac objects or radio galaxies.
We conclude that dust associated absorption represents a viable explanation 
for the observed effect.

In view of this result it seems surprising that we do not find any
difference in the X-ray detection probability for the red and the blue
stellar objects. Since the observed X-ray flux should be reduced by absorption
for 
the reddened sources, it would have been expected that also the fraction of
upper limits is higher among these sources. As it is shown in Section 5.2,
this is obviously not the case. However, in Section 4.1 we concluded that
distance is the most important factor, which determines the detection 
probability in soft X-rays and the absorption effect as discussed above
is not strong enough in order to dominate over distance. Indeed, for a given
intrinsic luminosity and reasonable amounts of cold gas, the observed soft 
X-ray flux is at most reduced by a factor of three due to absorption, 
whereas it declines already by a factor of $\sim$3.5, when the source is 
shifted from $z = 0.1$ to $z = 0.2$. 

Since many of the spectral parameters were determined from the hardness ratios,
which usually results in relatively large uncertainties, it is impossible
to derive any stringent constraints on the amount of dust associated absorption
from the X-ray spectra of the Parkes quasars. We find only six objects with a 
significant $\Delta N_{\rm H} = N_{\rm H,free} - N_{\rm H,gal} > 0$, 
however, all of them at high redshift. Thus, the implied amounts of dust 
(again assuming Galactic gas-to-dust ratios) would be much higher than 
observed in the optical spectra. This apparent discrepancy has already been
noted previously (e.g. Elvis et al. 1994) and several solutions have been
proposed. First of all, all estimations of the expected amount of X-ray
absorption (or optical reddening) are based on the assumption of Galactic
gas-to-dust ratios and it is not at all sure, whether this assumption is 
still valid at high redshifts. Indeed, Pei, Fall \& Bechtold (1991) show that
the gas-to-dust ratio is at least ten times higher in damped Ly$\alpha$ 
systems. Therefore a high neutral gas column must not necessarily imply a 
correspondingly large dust column and therefore significant X-ray absorption 
must not be accompanied by a large optical extinction. Similarly, the metal 
abundances could be reduced at large redshifts, which also is indicated by the 
properties of damped Ly$\alpha$ systems (e.g. Meyer \& Roth 1990; Turnshek 
et al. 1989). 
 
Finally, also the observed large scatter in the $\log (f_{\rm x}/f_{\rm o})$ 
distribution does not allow to draw any firm conclusions about the amount
of dust present in radio-loud quasars. In particular, the observed deficit of
high redshift sources with low $\log (f_{\rm x}/f_{\rm o})$ ratios is 
most likely not due to dust extinction, because the optical continua of the
high redshift sources are not particularly red. We note, however, that large 
observed scatter is also not inconsistent with the dust hypothesis as it was 
claimed previously based on the analysis of optically and X-ray selected 
samples.

One explanation for the limited evidence for dust associated absorption in
the X-ray data could be that the large spread in the $B_{\rm J} - K$ colors 
reported by Webster et al. (1995) is not due to dust. Many alternative
suggestions have been made to explain the observed $B_{\rm J} - K$ values,
such as an intrinsically red optical continuum, a host galaxy contribution 
\cite{benn} or relativistic beaming \cite{srianand}. It is beyond the scope
of this paper to discuss the possible origins of the observed $B_{\rm J} - K$
colors. This will be done in a future paper (Francis et al., in prep).
Here we only note that detailed investigations of the optical spectra of many 
of the red sources show that an intrinsic origin for the red colors gives the 
best fit to the spectra of around half of the red sources, but the remainder, 
including many of the reddest sources do show signs of dust reddening 
(Francis et al., in prep.). A similar result has been reported by Puchnarewicz \& Mason 
\shortcite{puchna}, who find evidence for reddening by dust among those RIXOS 
AGN with very red optical continua.

If one accepts the dust hypothesis, then why doesn't the neutral gas 
associated with the dust show up more clearly in the soft X-ray regime?
Various explanations are conceivable. First of all, it has already been
noted that all estimations are based on the assumption that the gas-to-dust
ratio in all quasars is the same as that in our Galaxy. If the gas-to-dust
ratio is lower, the observed dust extinction is not accompanied by a large
neutral gas column and the corresponding X-ray absorption might not be
measurable given the limited sensitivity of the presented data. Secondly,
the ionization state of the gas associated with the dust is unknown. If 
the gas is ionized, its X-ray opacity is reduced and it would therefore allow 
for larger amounts of dust without strongly affecting the soft X-ray emission.
So-called 'dusty warm absorbers' have recently been claimed to exist in a 
couple of sources, e.g. IRAS 1334+2438 \cite{brandt} and NGC 3786 
\cite{komossa}. Clearly, the quality of the available X-ray spectra of the
sources presented in this paper is not sufficient to constrain the ionization 
state of the absorbing gas.  

High quality X-ray spectra are needed to directly quantify the amount dust 
probably present in optically red quasars by measuring the carbon absorption 
edge at $\sim 0.3$ keV. These are currently not available, but future X-ray 
missions like XMM and AXAF, which combine a large effective area with high 
energy resolution down 0.2 keV, should be able to provide the necessary data.

\section{Conclusions}

Using {\em ROSAT} All--Sky Survey data and pointed PSPC observations we
determined the X-ray properties of all 323 objects from the Parkes sample 
of flat-spectrum radio sources as defined in Drinkwater et al. 
\shortcite{drinkwater}.
The results are as follows:
\begin{enumerate}
\item 163 sources were detected at the 3$\sigma$ level. For the remaining
160 sources 2$\sigma$ upper limits to the soft X-ray flux were determined. 
\item Using a hardness ratio technique and explicit fits to the data we 
determined the soft X-ray spectra of 115 sources. The average power-law 
photon index for flat-spectrum quasars is $\langle\Gamma\rangle = 1.95\pm 
0.13$, slightly flatter ($\Delta\Gamma \sim 0.2$) than for 'ordinary' 
radio-loud quasars. The average photon index for BL Lac objects is 
$\langle\Gamma\rangle = 2.40^{+0.12}_{-0.31}$, whereas radio galaxies 
generally display harder soft X-ray spectra ($\langle\Gamma\rangle = 1.70\pm 
0.23$).
\item We confirm the inverse correlation of the spectral index with redshift
for radio-loud quasars. We also find a significant inverse correlation of 
the photon index with the radio spectral index between 2.7\,GHz and 5\,GHz,
which is in accord with current orientation dependent unification schemes
for radio-loud AGN.
\item Correlations of X-ray with total radio luminosity were found for 
quasars, galaxies and BL Lac objects. Partial correlation analyses indicate
that a redshift effect is negligible for quasars and galaxies, whereas it
might influence the correlation for BL Lacs.
\end{enumerate}

The question, whether intrinsic dust is the origin of the observed red optical
continua of many of the quasars of the sample, cannot be unambiguously 
answered on the basis of the presented X-ray properties. Nevertheless, it is 
tempting to interpret the redshift dependence of the difference in the 
observed X-ray luminosities for optically red and blue sources in terms 
of dust associated absorption. However, firm conclusion on the basis of current
X-ray data are hampered by the fact that all arguments rely on implicit
assumptions, which may not be valid in general, such as Galactic gas-to-dust
ratios in quasars and a low ionization state of the X-ray absorbing gas
associated with dust.

\section*{Acknowledgments}
The {\em ROSAT} project is supported by the Bundesministerium f\"ur
Forschung and Technologie (BMBF). We thank our colleagues from the
{\em ROSAT} group for their support.
This research has made use of the NASA/IPAC Extragalactic Data Base
(NED) which is operated by the Jet Propulsion Laboratory, California
Institute of Technology, under contract with the National Aeronautics
and Space Administration.

\end{document}